\documentclass[preprint,12pt]{elsarticle}

\usepackage{amssymb}
\journal{Nuclear Physics B}

\begin{document}

\begin{frontmatter}



%\title{Reconstruction of the Neutrino Factory Golden Channel in a Magnetised Iron Detector}
%\title{Performance of the MIND detector at a Neutrino Factory}
%\title{Efficient wrong-sign muon reconstruction in the Neutrino Factory MIND}

\title{Performance of the MIND detector at a Neutrino Factory using realistic muon reconstruction}


\author[label1]{A. Cervera}
\ead{anselmo.cervera@ific.uv.es}
\author[label2]{A. Laing\corref{cor1}}
\ead{a.laing@physics.gla.ac.uk}
\author[label1]{J. Mart\'\i n-Albo}
\ead{Justo.Martin-Albo@ific.uv.es}
\author[label2]{F.J.P. Soler}
\ead{p.soler@physics.gla.ac.uk}

\cortext[cor1]{Corresponding author}

\address[label1]{Instituto de F\'\i sica Corpuscular, CSIC $\&$ Universidad de Valencia, Valencia, Spain}

\address[label2]{Department of Physics $\&$ Astronomy, University of Glasgow, Glasgow, U.K.}

\begin{abstract}
A Neutrino Factory producing an intense beam composed of $\nu_e (\overline{\nu}_e)$ and $\overline{\nu}_\mu (\nu_\mu)$ from muon decays has been shown to have the greatest sensitivity to the two currently unmeasured neutrino mixing parameters, $\theta_{13}$ and $\delta_{CP}$. Using the `wrong-sign muon' signal to measure $\nu_e \rightarrow \nu_\mu (\overline{\nu}_e \rightarrow \overline{\nu}_\mu)$ oscillations in a 50~ktonne Magnetised Iron Neutrino Detector (MIND) sensitivity to $\delta_{CP}$ could be maintained down to small values of $\theta_{13}$. However, the detector efficiencies used in 
%these 
previous studies were calculated assuming perfect pattern recognition. In this paper, MIND is re-assessed taking into account, for the first time,
% muon charge current event pattern recognition and Kalman filter reconstruction. 
a realistic pattern recognition for the muon candidate. 
Re-optimisation of the analysis utilises a combination of methods, including a multivariate analysis similar to the one used in MINOS, to maintain high efficiency while suppressing backgrounds, ensuring that the signal selection efficiency and the background levels are comparable or better than the ones in previous analyses. 
%As a result MIND remains the most sensitive future facility for the discovery of CP violation from neutrino oscillations.

\end{abstract}


\end{frontmatter}


\section{Introduction}
%**********************************************************************
\label{Intro}
The concept of a neutrino beam from the decay of muons in a storage ring was first proposed in 1980 \cite{Cline:1980sa}. More recently, such a facility was explored as a preliminary stage towards a muon collider and was renamed ``Neutrino Factory". Its physics potential was originally described by Geer \cite{Geer:1997iz}. The great advantage of a Neutrino Factory over conventional neutrino beams from pion decay is that the decay of muons is very well described by the Standard Model and so the beam flux is easily calculable. Therefore, it is possible to perform high precision neutrino oscillation experiments at a high flux Neutrino Factory. Another significant feature of a Neutrino Factory is that one can accelerate muons of both signs into a storage ring, thereby enabling study of both neutrino and anti-neutrino oscillations with equal flux, vastly improving sensitivity to CP violation in the neutrino sector. For a more recent review see~\cite{Geer:2009zz}.

Early papers on the physics outcomes of a Neutrino Factory concentrated on the sub-dominant $\nu_e \rightarrow \nu_\mu$ oscillation \cite{DeRujula:1998hd} in which a muon of opposite charge to that stored in the facility storage ring (wrong-sign muon) would be produced in a far detector by the charge current (CC) interactions of the oscillated $\nu_\mu$. The first analysis of the capabilities of a large magnetised iron detector to detect the wrong-sign muon signature was discussed in~\cite{Cervera:2000kp} (termed the Golden Channel), where it was demonstrated that this combination was capable of the extraction of the remaining unknown parameters in the neutrino sector, the third mixing angle $\theta_{13}$ of the Pontecorvo-Maki-Nakagawa-Sakata (PMNS) matrix \cite{Pontecorvo:1957cp,Pontecorvo:1957qd,Maki:1962mu} and the $CP$ violating phase $\delta_{CP}$. 

The Magnetised Iron Neutrino Detector (MIND) is a large scale iron and scintillator sampling calorimeter. As a result of the studies mentioned above it is considered the baseline detector for a Neutrino Factory (NF) storing muons in the energy range 20-50~GeV~\cite{Abe:2007bi}. Under the remit of EUROnu \cite{EUROnu} and the International Design Study for a Neutrino Factory \cite{IDS-NF} all aspects of possible future neutrino beam facilities including accelerator, detectors and physics must be studied and compared to select the best option to determine the remaining oscillation parameters. 

Previous studies of MIND focused on the topology and kinematics of neutrino events in the detector, assuming perfect pattern recognition.  
%of the $\nu_\mu$ CC interactions. %Anselmo
By smearing the kinematic variables of the participant muon and hadronic shower it was demonstrated that using a combination of cuts on the relative length of the two longest particles in the event and the momentum and isolation of this candidate, high signal identification efficiency and background suppression could be achieved~\cite{Cervera:2000vy,CerveraVillanueva:2008zz}. However, a full study without such assumptions is necessary to fully characterise the detector response.

While MIND is essentially a large scale version of the MINOS detector~\cite{michael-2008-596}, the nature of the NF beam -- containing 50\% $\nu_e$ and 50\% $\overline{\nu}_\mu$ in the case of stored $\mu^+$ -- means that the optimisation of the analysis is somewhat different. Incorrect charge assignment (charge misidentification) of 
%non-oscillation %Anselmo 
non-oscillated 
$\overline{\nu}_\mu$ CC interactions present a significant possible background in this beam configuation, in addition to backgrounds from meson decays in the hadronic shower and misidentification of Neutral Current (NC) and $\nu_e$ CC events.

This current study re-visits the problem by taking an un-biased look at the visible part of a large sample of neutrino interactions -- generated using the same GEANT3~\cite{geant3wu} simulation as in the above mentioned studies with a uniform distribution in neutrino energy -- and developing pattern recognition algorithms (first presented in~\cite{Cervera:2008nf}) -- described in Sec.~\ref{rectool} -- to extract a candidate muon for fitting using a Kalman filter. Successful fits are then subject to offline analyses -- described in Sec.~\ref{OffAn} -- to determine the validity of those wrong sign candidates. Analysis results are presented in Sec~\ref{res}.
%Anselmo Quick decription of the paper sections missing. You only mentioned one section

%**********************************************************************
\section{MIND parameterisation and expected event yields}%maybe separate section altogether?
%**********************************************************************
\label{MINpar}
For the purpose of the described study, MIND is a cuboidal detector of $14\mbox{ m}\times14$~m cross-section and $40$~m length, segmented as $4$~cm of iron and $1$~cm of plastic scintillator for a total mass of $\sim$51.0 ktonnes. A dipole magnetic field of mean induction $1$~T in the transverse plane provides the field neccessary for charge 
and momentum %Anselmo
measurements.

%In Sec.~\ref{res}, 
In the first part of the analysis, event vertices were generated centred in the detector plane at $1.5$~m from the front of the detector in the beam direction ($z$) in order to study the nature of the backgrounds without detector edge effects. Sec.~\ref{sec:fidcut} discusses the expected fiducial effects when a more realistic randomly generated vertex is considered.

At a MIND placed $4000$~km from the neutrino source and assuming the current best global fit oscillation parameters: $\theta_{12} = 33.5^{\circ} \mbox{, } \theta_{13} = 5.7^{\circ} \mbox{, } \theta_{23} = 45^{\circ} \mbox{, } \Delta m^2_{21} = 7.65\times10^{-5} \mbox{~eV}^2\mbox{, } \Delta m^2_{32} = 2.40\times10^{-3} \mbox{~eV}^2$~\cite{Schwetz:2008er}, setting $\delta_{CP} = 45^{\circ}$ and calculating matter effects using the PREM model~\cite{Dziewonski:1981xy}, the expected total number of interactions due to $10^{21}$ $\mu^+$ decays at $25$~GeV energy would be of order those shown in table~\ref{tab:int}.

\begin{table}[ht]
  \begin{center}
    \begin{tabular}{|c|c|c|c|}
      \hline
      $\overline{\nu}_\mu$ CC & $\nu_e$ CC & $\overline{\nu}_\mu + \nu_e$ NC & $\nu_\mu$ (Signal) \\
      \hline
      $1.22\times10^5$ & $3.34\times10^5$ & $1.48\times10^5$ & $5.56\times10^3$\\
      \hline
    \end{tabular}
  \end{center}
  \caption{\emph{Expected absolute number of interactions in a 51 ktonne MIND at a distance of 4000~km from a NF storage ring with 25~GeV muons.}}
  \label{tab:int}
\end{table}

Thus in order to successfully extract oscillation parameters from the golden channel, potential backgrounds from non-signal interactions must be suppressed to at most the $10^{-3}$ level in absolute terms. Moreover, the existence of possible degenerate solutions due to uncertainty in the measured parameters and due to the nature of the oscillation probability (see~\cite{BurguetCastell:2002qx,Barger:2001yr}) means that spectral information is required to reliably determine $\delta_{CP}$. This additional requirement dictates that backgrounds must be suppressed to below $10^{-3}$ in each energy bin while maintaining an efficiency threshold below 5~GeV so that information on the rise of the 
%second %Anselmo 
first 
oscillation maximum is available.%the requirement of energy binning information to determine $\delta_{CP}$ reliably means that the background levels must be suppressed to a mean level of $10^{-3}$ in each bin while maintaining an efficiency threshold below $5$~GeV.%more and better description

%**********************************************************************
\section{Reconstruction tools}
%**********************************************************************
\label{rectool}
The reconstruction package was used to analyse a large data set comprised of Deep Inelastic Scattering (DIS) neutrino interactions of $\overline{\nu}_\mu$ and $\nu_e$ generated by the LEPTO61~\cite{Ingelman:1997Cp} package and tracked through the GEANT3 simulation of MIND. Considering CC interactions of $\overline{\nu}_\mu$ and $\nu_e$ with a dedicated study of events containing wrong sign muons from meson decay in $\nu_\mu$ CC and NC interactions the main expected backgrounds were studied.

Each event considered comprised all three dimensional points with their associated energy deposit, which were recorded in the scintillator sections of the MIND simulation, with the \emph{x,y} position of these hits smeared according to a $\sigma = 1$~cm Gaussian before analysis began.

%--------------------------------------------------------------
\subsection{Muon candidate extraction}
%--------------------------------------------------------------
\label{MuCan}
After ordering the hits from smallest to greatest $z$ position in the detector the first step of the reconstruction 
was to extract a candidate muon from the event. Two methods were employed to perform this task depending on the event topology:  
a Kalman filter incremental fit was used to extract candidates from those events with one particle clearly longer than the others 
(described in Sec.~\ref{KalPat}), while a Cellular Automaton method was used in those events not viable for reconstruction  
through the first method (see Sec.~\ref{cellPat}). The criterion for the first category was that the five planes with activity furthest downstream should contain no more than one hit per plane.

%--------------------------------------------------------------
\subsubsection{Kalman Filter candidate extraction}
%--------------------------------------------------------------
\label{KalPat}

\begin{figure}
  \begin{center}
    \includegraphics[scale=0.33]{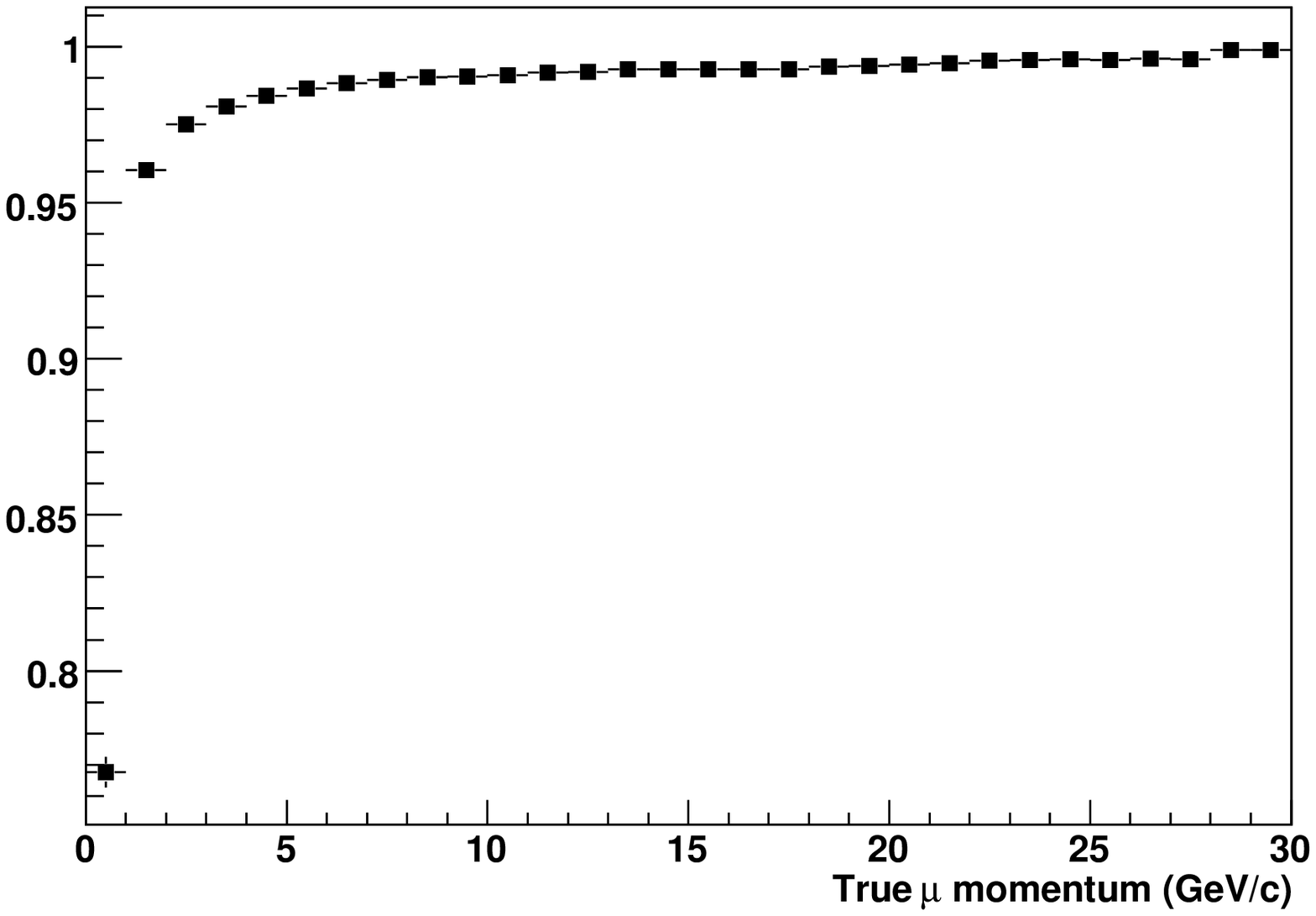} 
    \includegraphics[scale=0.33]{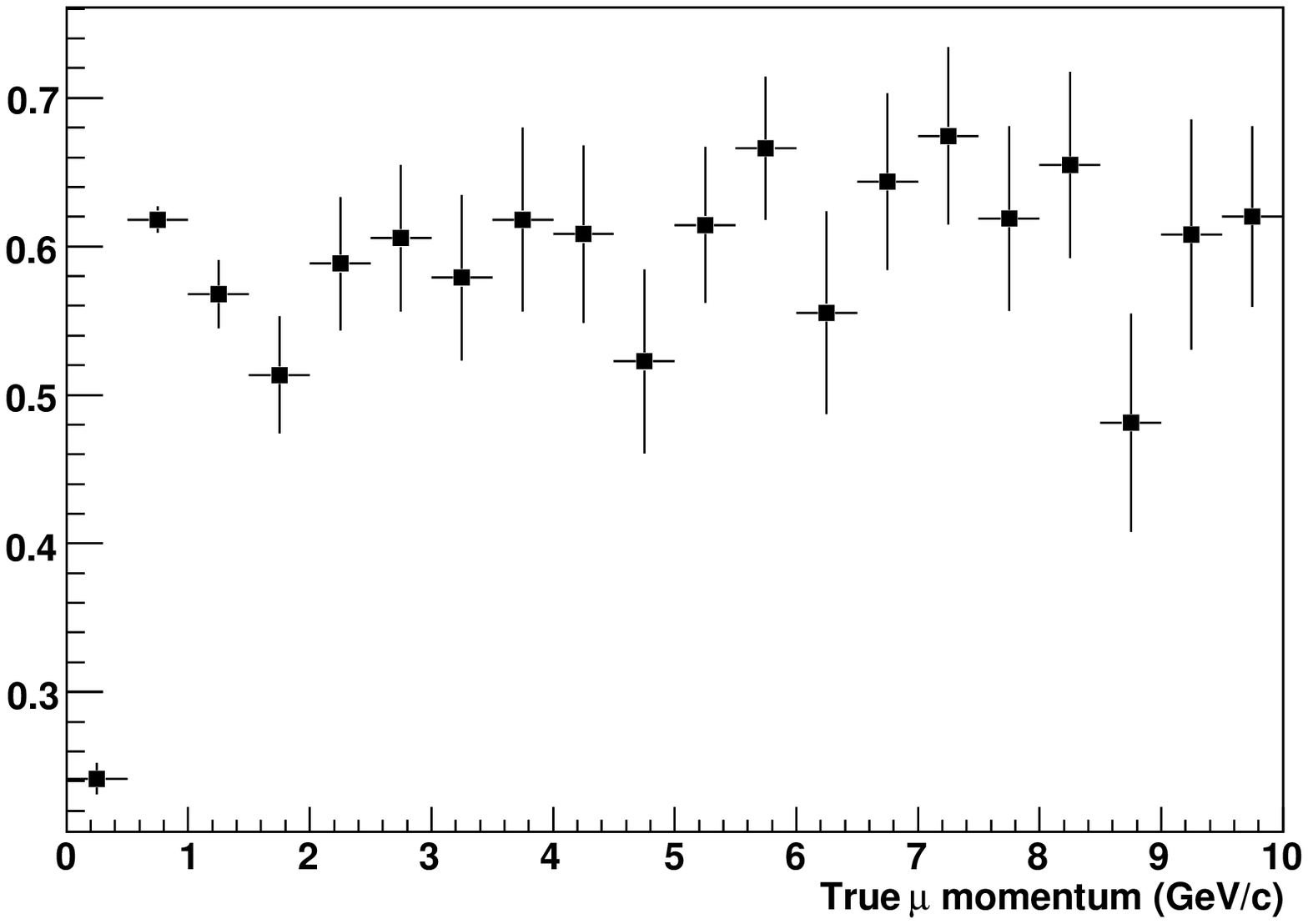}
  \end{center}
  \caption{\emph{Muon candidate purity as a function of true muon momentum for i) the Kalman filter extraction method and ii) the Cellular automaton method.}}
  \label{fig:methPur}
\end{figure}

Using the Kalman filter algorithm provided by RecPack~\cite{CerveraVillanueva:2004kt} 
it is possible to propagate the track parameters 
back through the planes using a helix model, which takes into account multiple scattering and energy loss. 
Since, in general, a muon will act as a Minimum Ionising Particle (MIP) 
and will travel further in the detector than the hadronic particles, 
those hits furthest downstream can be assumed to be muon hits and used as a seed for the Kalman filter. 
The seed state is then propagated back to each plane with multiple hits and the matching $\chi^2$ to each of the hits is computed. 
Hits with matching $\chi^2$ below 20 are considered and in each plane the one with the best matching among these is added to the trajectory and filtered (the track parameters 
are updated with the new information). All accepted hits constitute the candidate muon and are presented for fitting 
(Sec.~\ref{KalFit}), with the remaining hits being considered as hadronic activity. Fig.~\ref{fig:methPur}-(left) shows the fraction of true muon hits in the candidate when using this method.
%Maybe purity plot??
%--------------------------------------------------------------
\subsubsection{Cellular Automaton candidate extraction}
%--------------------------------------------------------------
\label{cellPat}
Events with high $Q^2$ transfer or low neutrino energy can tend to be rejected by the first method, 
since in general the muon will not escape the region of hadronic activity.   
%becouse there requirement of the minimum number of single occupancy planes is not fulfilled.   
%which do not pass the minimum `free' hit cut
In order to recover these events a second method is employed. The Cellular Automaton method 
(based on the method described in~\cite{Emeliyanov_otr/itr-cats:tracking}) 
uses a neighbourhood function to first rank all the hits and then form all viable combinations 
into possible trajectories.

A `neighbour' is defined as a hit in an adjacent plane within a pre-defined transverse distance of the projection into 
that plane of the straight line connecting hits in the previous two planes. Starting from the plane with lowest $z$ position,
 hits are given a rank one higher than their neighbour in the previous plane should they have one. Trajectories are then formed 
from every possible combination of one hit per plane starting with those of highest rank using the neighbourhood 
function with a stricter condition.

Those trajectories formed using this method are then subject to a number of tests to determine which is most likely 
to be a muon. After having a basic helix fit performed and being assessed according to their length, trajectories 
are rejected for being short, having high $\chi^2$ fit or high relative curvature error (described in Sec.~\ref{qualcut}). 
The candidate muon is then selected as the longest remaining trajectory with the lowest $\chi^2$. All other hits in the 
event are considered to be from hadronic activity. Fig.~\ref{fig:methPur}-(right) shows the purity of the candidate when using this method.

%--------------------------------------------------------------
\subsection{Candidate fitting}
%--------------------------------------------------------------
\label{KalFit}
All candidates successfully extracted from their event that have greater than six hits are presented to the fitter as a candidate muon. 
The same Kalman filter algorithm is used here as in Sec.~\ref{KalPat}. Fitting the candidate iteratively improves seeding and thus using 
a more constricted $\chi^2$ condition than in the pattern recognition, the maximum number of successful, reliable fits were achieved.

With the trajectory hits ordered in increasing $z$ position, a least squares quartic fit was performed on the 
section outside the planes where there was hadronic activity. This fit was used to estimate the slopes in $x$ and $y$ 
and the momentum of the candidate, to be used as a seed for the Kalman filter helix fit in the forward direction.   
The matching $\chi^2$ was once again checked at each hit. Hits with greater than 
the pre-determined maximum (20) were ignored. In addition, the filtering process only allows a pre-determined maximum number 
of hits (5) to be ignored. Should this number be reached, the filtering process is aborted and the smoother uses only those 
hits up to this point in the candidate. This method efficiently rejects hits beyond any large angle scatter which could 
cause charge misidentification. Successful fits were re-seeded with the state vector at the first fitted hit and a scalar 
multiple (5) of the corresponding covariance matrix (taking only the diagonal elements) and then refitted.

Failed fits and those with less than 50\% of their hits fitted are then fitted again in the backwards direction using 
the seed from the pattern recognition. Two iterations are once again performed, with successful fits being accepted 
and those which are unsuccessful reverting to the result of the original fit.

The result of a fit being the track parameters at the projection to the event true vertex \emph{z} position (3-momenta, position and charge).%Anselmo

%--------------------------------------------------------------
\subsection{Hadronic reconstruction}
%--------------------------------------------------------------
\label{HadFit}
The hadronic activity must be used to reconstruct the energy of the hadronic shower 
in order to ultimately reconstruct the energy of the interacting neutrino. 
In the absence of a well developed algorithm to perform this task, the current study 
assumes reconstruction of the hadronic energy $E_{had}$ with a resolution $\delta E_{had}$ 
equal to that recorded by the MINOS CalDet testbeam~\cite{michael-2008-596,Adamson:2006xv}:
\begin{equation}
  \label{CalDet1}
  \frac{\delta E_{had}}{E_{had}} = \frac{0.55}{\sqrt{E_{had}}} \oplus 0.03.
\end{equation}

It was demonstrated in~\cite{Cervera:2000kp} that a cut based on the isolation of the muon candidate from the hadronic 
shower was a powerful handle for the rejection of hadronic backgrounds. %Anselmo
This isolation was measured via the $Q_t$ variable: 
\begin{equation}
  \label{eq:qt}
  Q_t = P_\mu \sin^2\vartheta,
\end{equation}
where $P_\mu$ is the muon momentum and $\vartheta$ is the angle between the muon 
and the resultant hadronic vector. This requires the reconstruction of the direction vector of the shower.  
%in order to gauge the candidate isolation. 
The Monolith test-beam~\cite{Bari:2003bt} measured an hadronic angular resolution described by:
\begin{equation}
  \label{eq:MONang}
  \delta \theta_{had} = \frac{10.4}{\sqrt{E_{had}}} \oplus \frac{10.1}{E_{had}}
\end{equation}
for a similar detector. This parameterization was used to smear the hadrom shower direction vector, 
which in combination with the reconstructed muon momentum and direction (see Sec.~\ref{KalFit}) 
were use to compute the $Q_t$ variable defined above. 

%%%%%Thus by smearing the true shower z angle, $\theta_{had}$, and reconstructing a smeared direction vector for the shower, the candidate-shower angle, $\vartheta$, can be used to gauge the isolatio%n of the candidate via the $Q_t$ parameter (described in Sec.~\ref{PQT}).

%--------------------------------------------------------------
%\subsection{Summary of reconstruction cuts}
%--------------------------------------------------------------

%\begin{itemize}
%\item Candidate extraction: Nisolated $\geqslant 5 \Rightarrow$ Kalman,   
%  \begin{itemize}
%  \item Kalman: no more that 3 skipped planes, good seed
%  \item CAT: chi2/ndf$<$20
%  \end{itemize}
%\item Fitting: successful fit
%\end{itemize}

%**********************************************************************
\section{Analysis tools and cuts}
%**********************************************************************

\label{OffAn}
%Successful fits are subject to a number of offline 
%analyses 
%in order to further reject possible background. 
As mentioned in Sec.~\ref{Intro} there are four main possible sources of background to the wrong sign muon search: 
incorrect charge assignment and high energy wrong sign muons from meson decays in $\overline{\nu}_\mu$ CC events, and NC and $\nu_e$ CC events wrongly identified as $\nu_\mu$ CC. 
In order to reduce these backgrounds while maintaining good efficiency a number of 
offline cuts were employed. They can be organised in four categories: i) muon candidate quality cuts, ii) $\nu_\mu$ CC selection cuts, 
iii) fiducial cuts and iv) kinematic cuts.

%--------------------------------------------------------------
\subsection{Muon candidate quality cuts}
\begin{figure}
  \begin{center}$
    \begin{array}{cc}
      \includegraphics[scale=0.37]{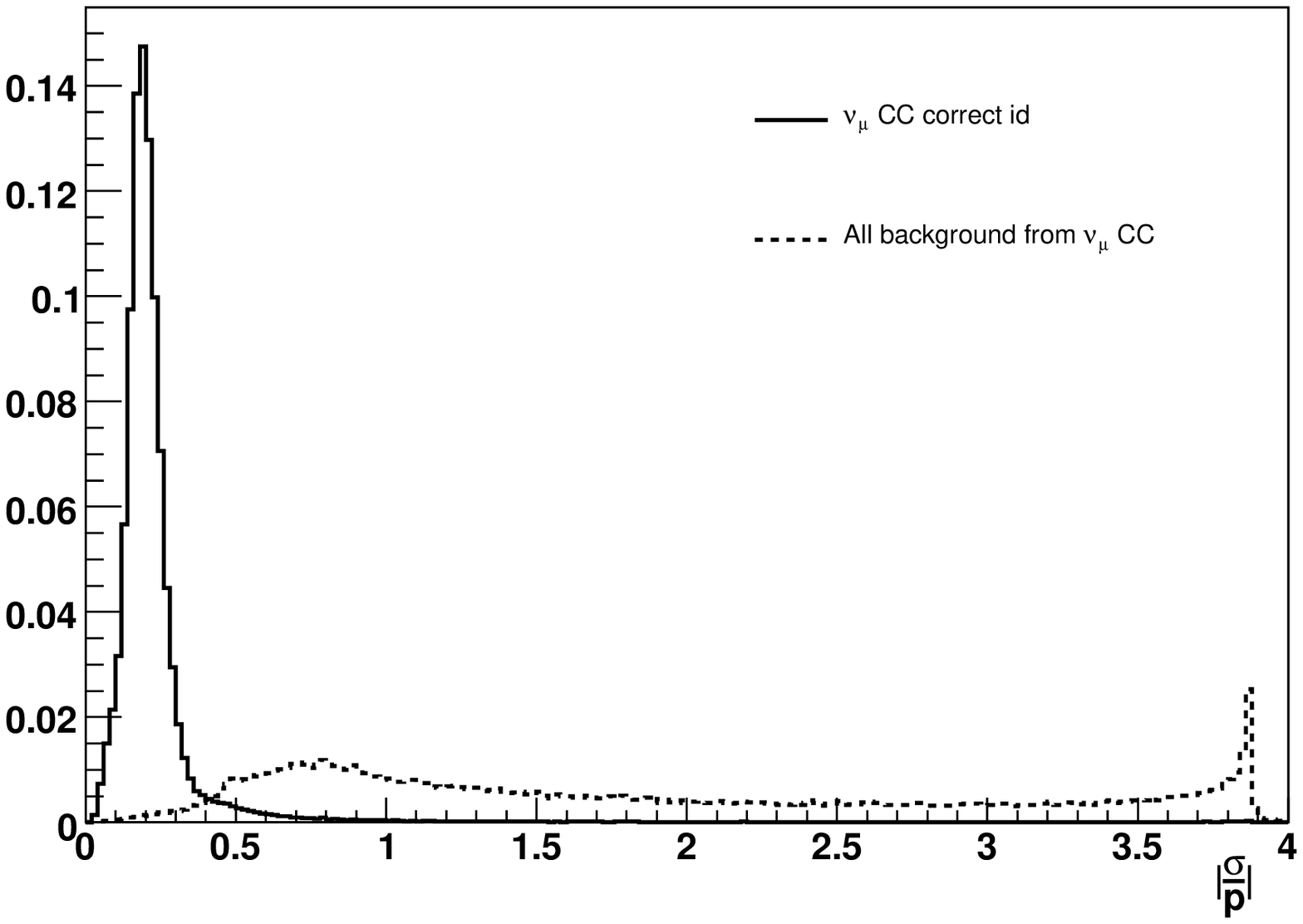} &
      \includegraphics[height=5cm,width=7cm]{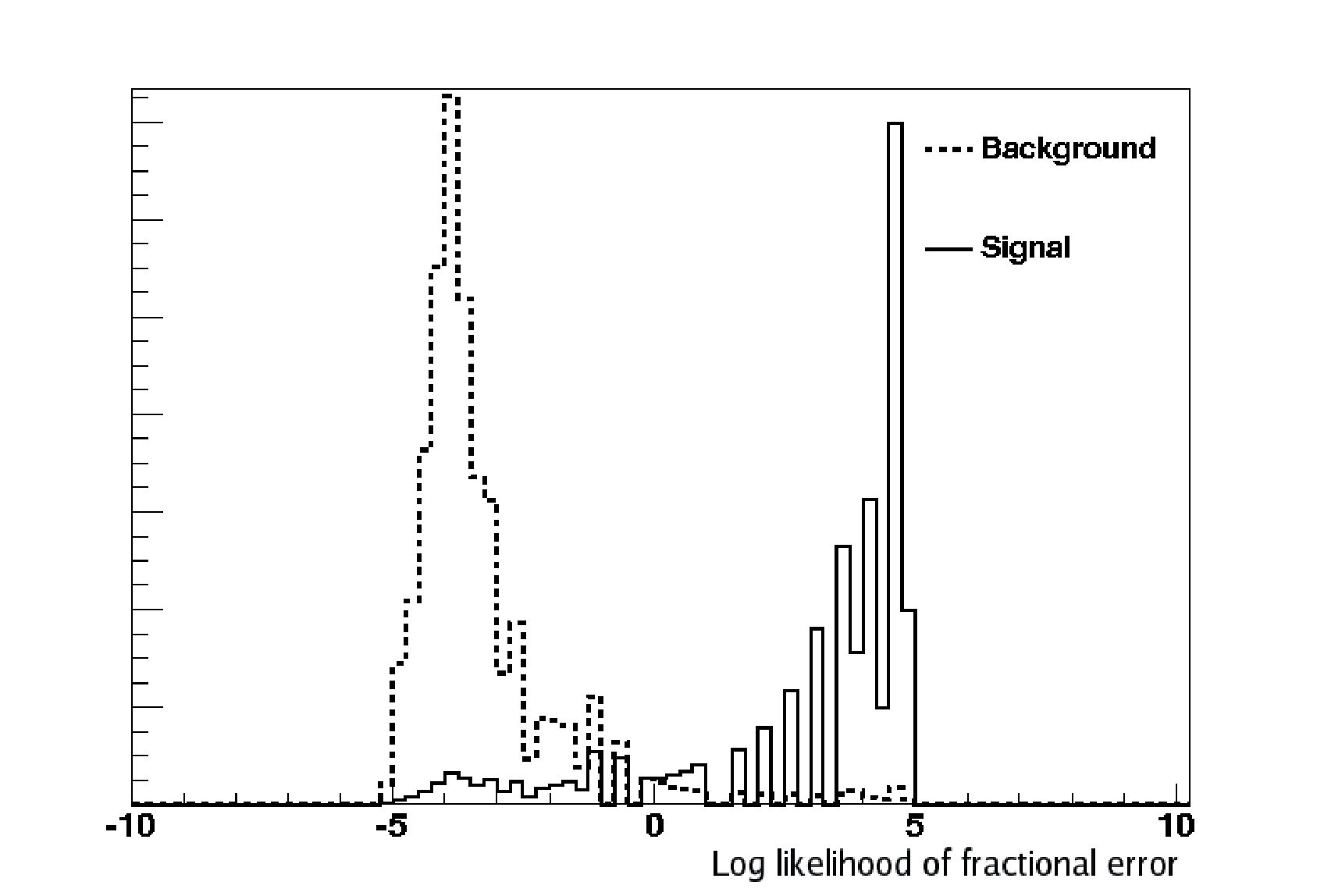}
    \end{array}$
  \end{center}
  \caption{\emph{i) $\displaystyle\frac{\sigma_{q/p}}{q/p}$ likelihood for 
signal and background from $\overline{\nu}_\mu$ CC, %Anselmo (Is this correct ? In the plot says All bkg from numucc)
      and ii) $\displaystyle\frac{\sigma_{q/p}}{q/p}$ log likelihood ratio ($\mathcal{L}_{q/p}$).}}
  \label{fig:errlike}
\end{figure}

%--------------------------------------------------------------
\label{qualcut}
These cuts are related to the quality of the candidate track fit and the determination of its
% momentum.  
curvature. %Anselmo
Two observables are considered: the $\chi^2$ probability of the Kalman filter fit and the relative error of the determined curvature 
($\frac{\sigma_{q/p}}{q/p}$). The first helps in rejecting high angle scatters or muon candidates with 
a large contamination from hadronic hits. The second variable is related with the probability of misidentifying the charge, 
and shows significant separation for correct and incorrect charge assignments as shown in Fig.~\ref{fig:errlike}-(left). 
%A candidate trajectory fitted with the incorrect charge will tend to be dominated by multiple scattering and hence the quality of the fit will be affected. 

It is possible to reject a large portion of possible backgrounds using sequential cuts on these two variables:
\begin{eqnarray}
  \label{eq:qualcut1}
  \left| \frac{\sigma_{q/p}}{q/p}\right| < 0.7 \mbox{ and } \chi^2_{prob} > 0.9999,
\end{eqnarray}
(where $\chi^2_{prob}$ is the $\chi^2$ probability as calculated in the TMath class of the ROOT framework~\cite{Brun:1997pa}).  
However, a slightly better rejection is found when the relative error cut is substituted 
by a cut on the log likelihood ratio of $\frac{\sigma_{q/p}}{q/p}$ for signal and background (see Fig.~\ref{fig:errlike}-(right)): 
\begin{eqnarray}
  \label{eq:qualcut2}
   \mathcal{L}_{q/p} > 2 \mbox{ and } \chi^2_{prob} > 0.9999.
\end{eqnarray}

%--------------------------------------------------------------
\subsection{$\nu_\mu$ CC selection cuts}
%--------------------------------------------------------------
\label{logLike}

The discrimination between $\nu_\mu$ CC and NC interactions relies on 
three easily available or calculable parameters, which are those of the extracted muon candidate.   
Due to the similarity of MIND and MINOS the parameters employed in the MINOS analysis~\cite{Adamson:2007gu} were used. 
Using a high statistic data set with knowledge of the true nature of each event, 
distributions of these three parameters for both NC and CC events were formed into PDFs (or likelihoods). 

\begin{figure}
  \begin{center}$
    \begin{array}{cc}
      \includegraphics[scale=0.35]{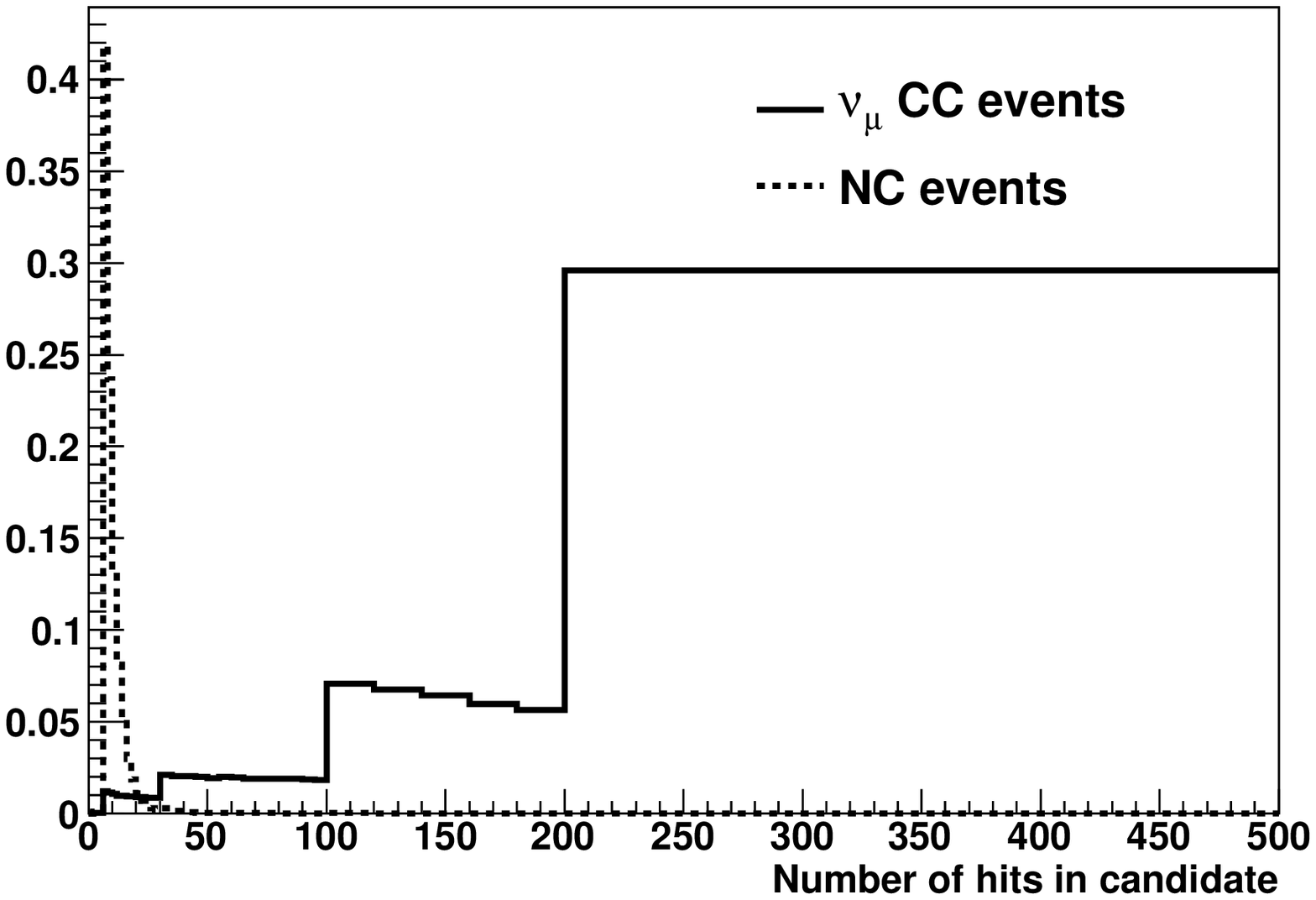} &
      \includegraphics[scale=0.35]{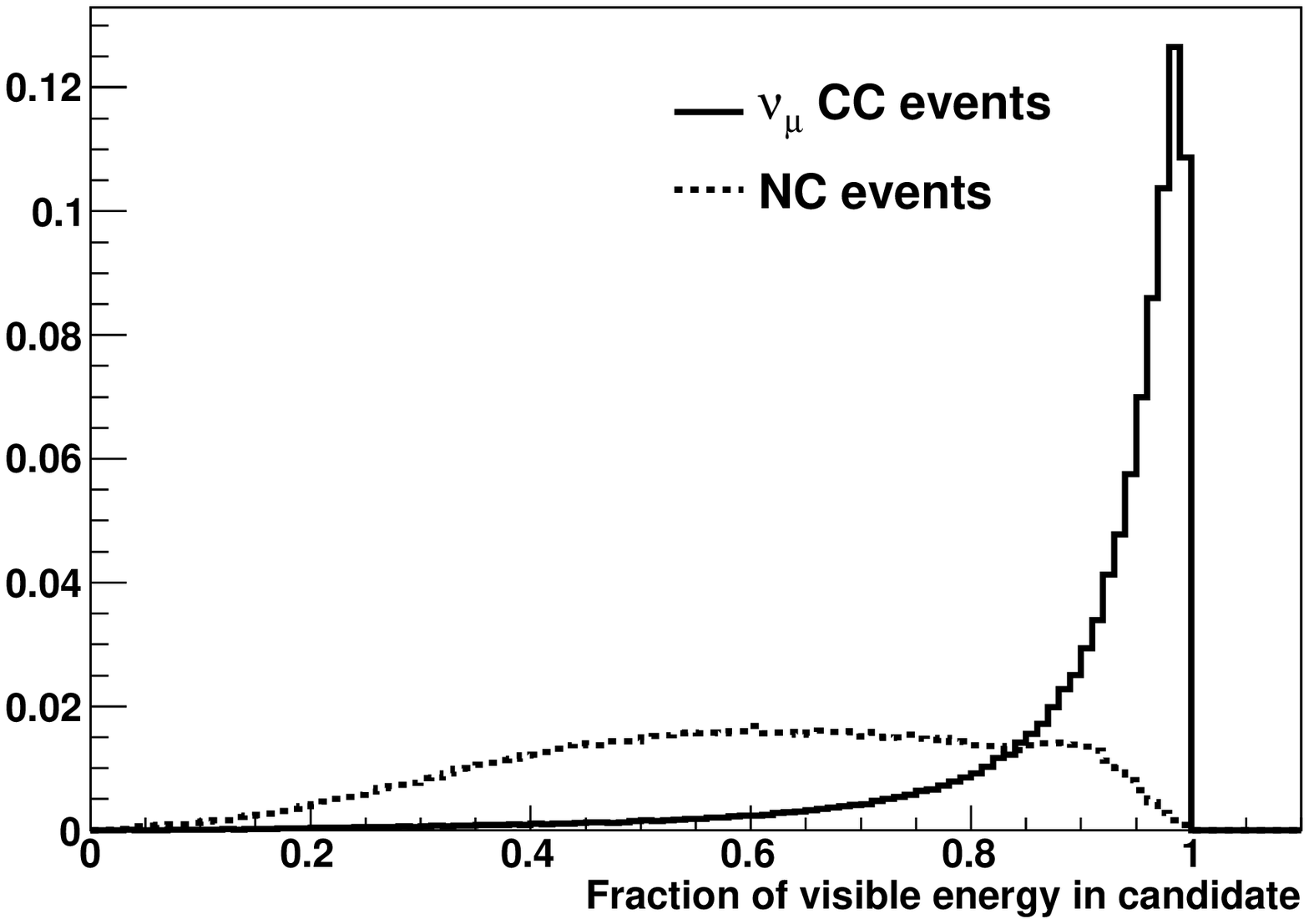}
    \end{array}$
      \includegraphics[scale=0.35]{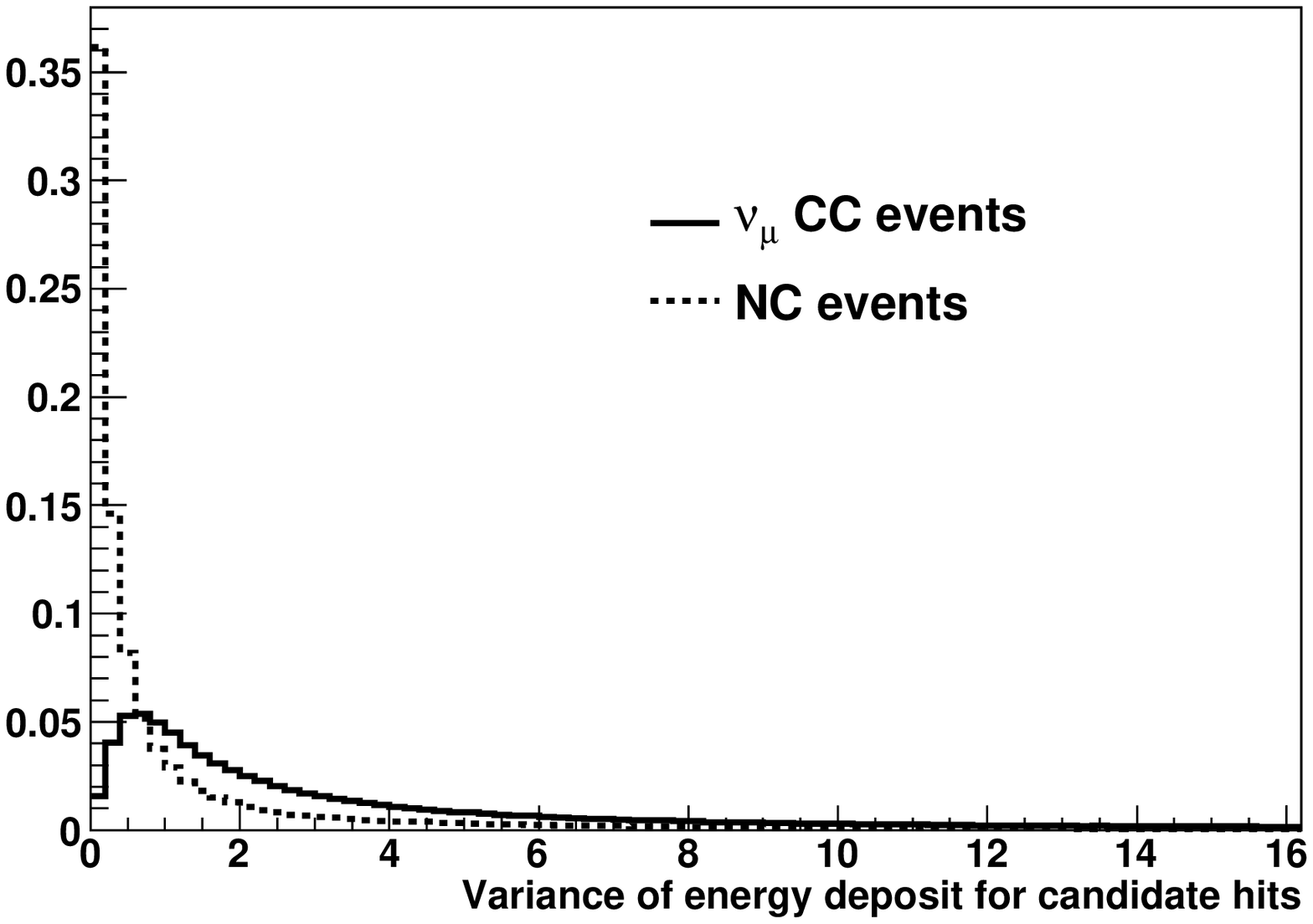}
  \end{center}
  \caption{\emph{PDFs of the three parameters used for NC/CC likelihood separation. i) Number of hits in candidate $l_{hit}$, ii) Fraction of visible energy in candidate $l_{frac}$  and iii) Variance of energy deposit in candidate $l_{var}$.}}
  \label{fig:pdfs}
\end{figure}

The first parameter was the length of the candidate in terms of the number of hits which form it ($l_{hit}$). 
This variable takes advantage of the nature of the muon as a penetrating particle and shows clear separation between 
$\nu_\mu$ CC and NC events (see Fig.~\ref{fig:pdfs}-(top-left)).
%While it is true 
%that candidates extracted from NC events tend to be much shorter than 
%those from $\nu_\mu$ CC (see Fig.~\ref{fig:pdfs}) the parameter is muon 
%momentum dependent and thus used alone would introduce an energy bias in the analysis.

The second parameter is the fraction of the total visible energy in the event which is in the candidate ($l_{frac}$). 
This parameter is not useful for all events due to the high probability for both NC and CC events to have a fraction 
very close to or equal to one. Thus, events that fall into this category, low $Q^2$ CC events or single pion production 
NC predominantly, are excluded from this distribution and do not use this parameter in their analysis. Here, while NC 
events demonstrate the full spectrum of possible values, signal events tend to be more concentrated at high fractions 
(see Fig.~\ref{fig:pdfs}-(top-right)). However, high $Q^2$ CC events will tend to exhibit NC like behaviour.

While the third parameter used by MINOS is the mean energy deposited per plane for the candidate, the current simulation setup of 
MIND does not exhibit sufficient separation in this parameter for effective analysis. 
Thus, in place of this parameter the variance of the deposit is used ($l_{var}$), shown in  Fig.~\ref{fig:pdfs}-(bottom).

\begin{figure}
  \begin{center}$
    \begin{array}{cc}
      \includegraphics[scale=0.35]{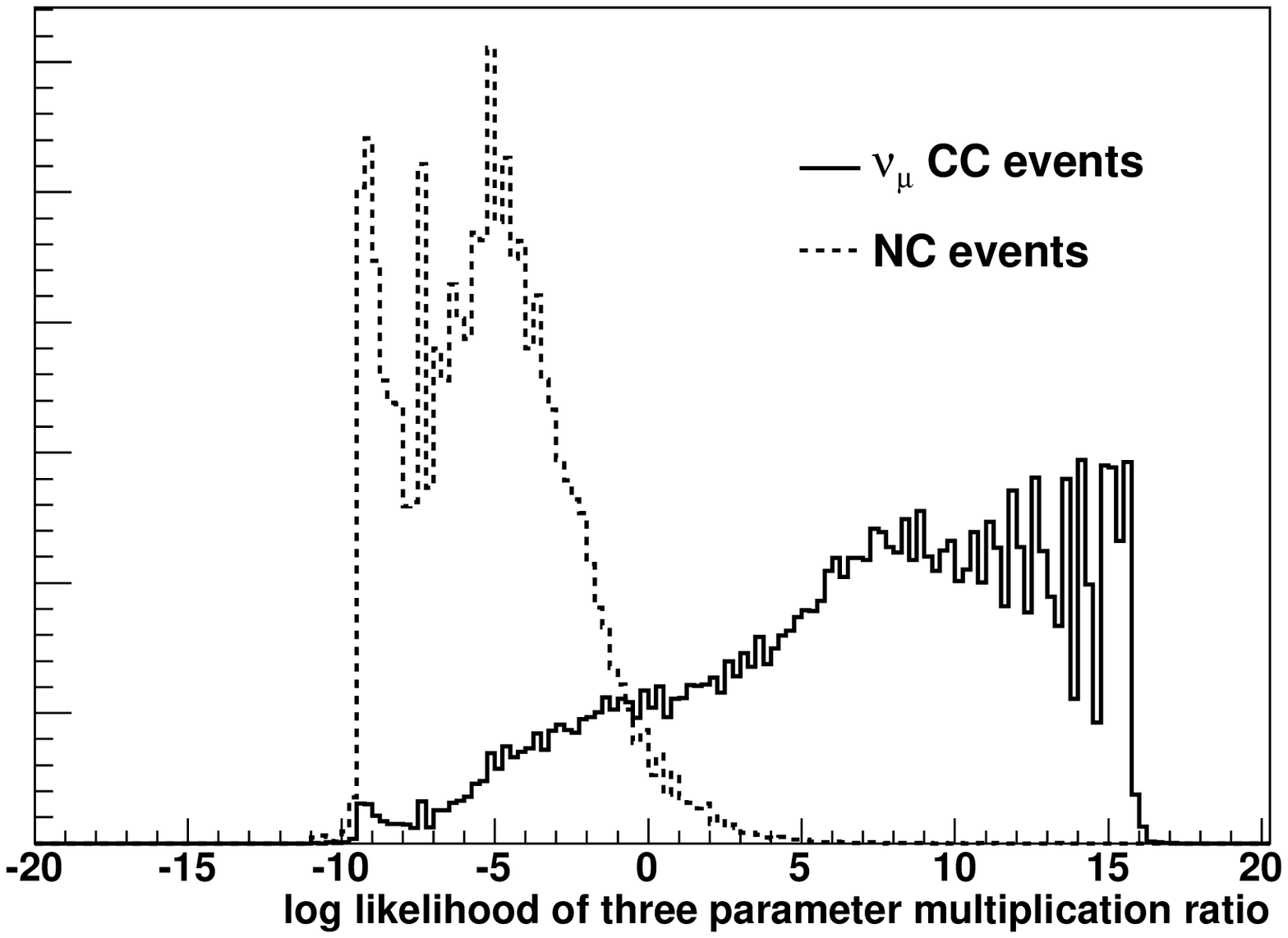} &
      \includegraphics[scale=0.35]{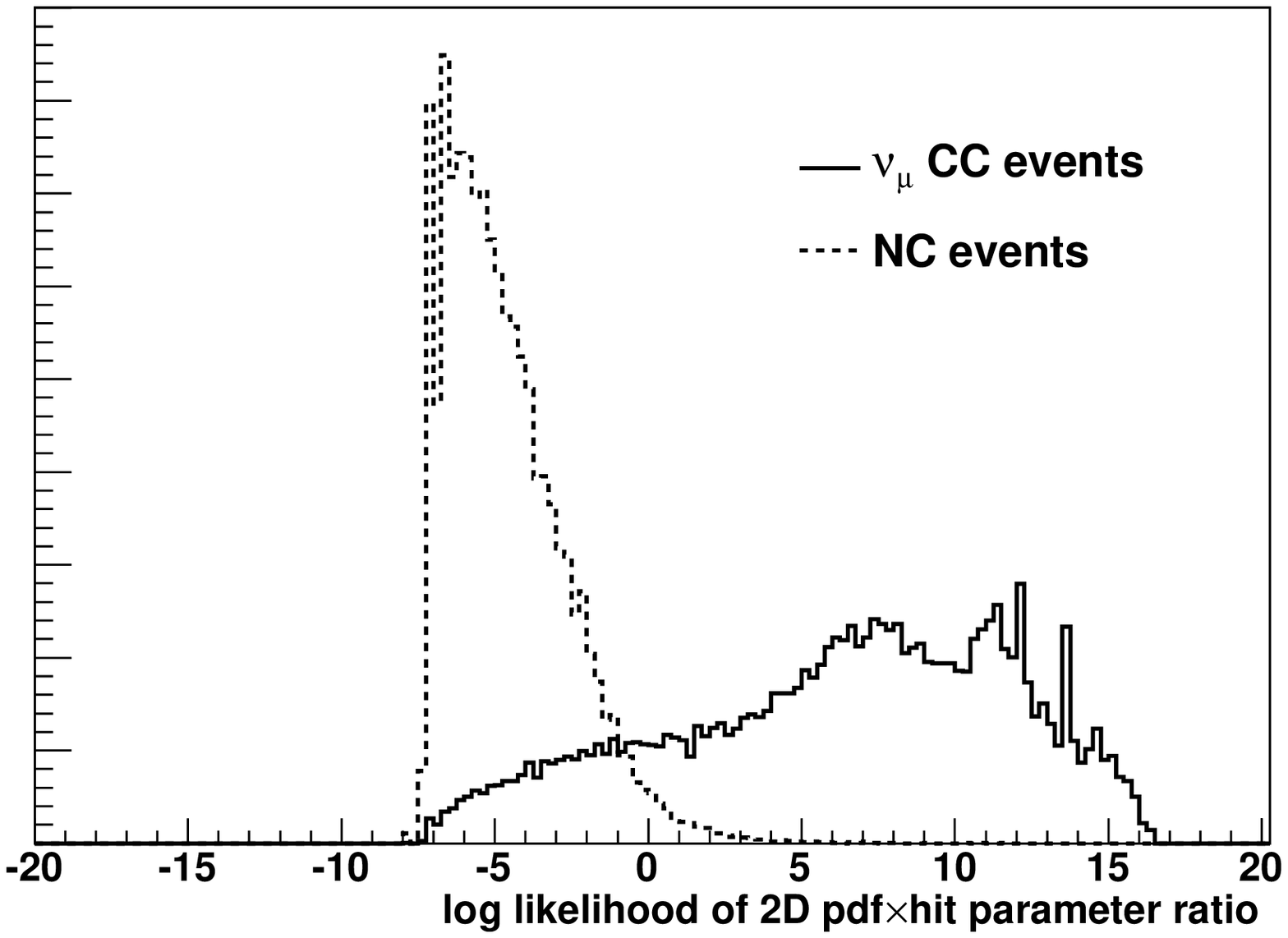} 
    \end{array}$
    \includegraphics[scale=0.35]{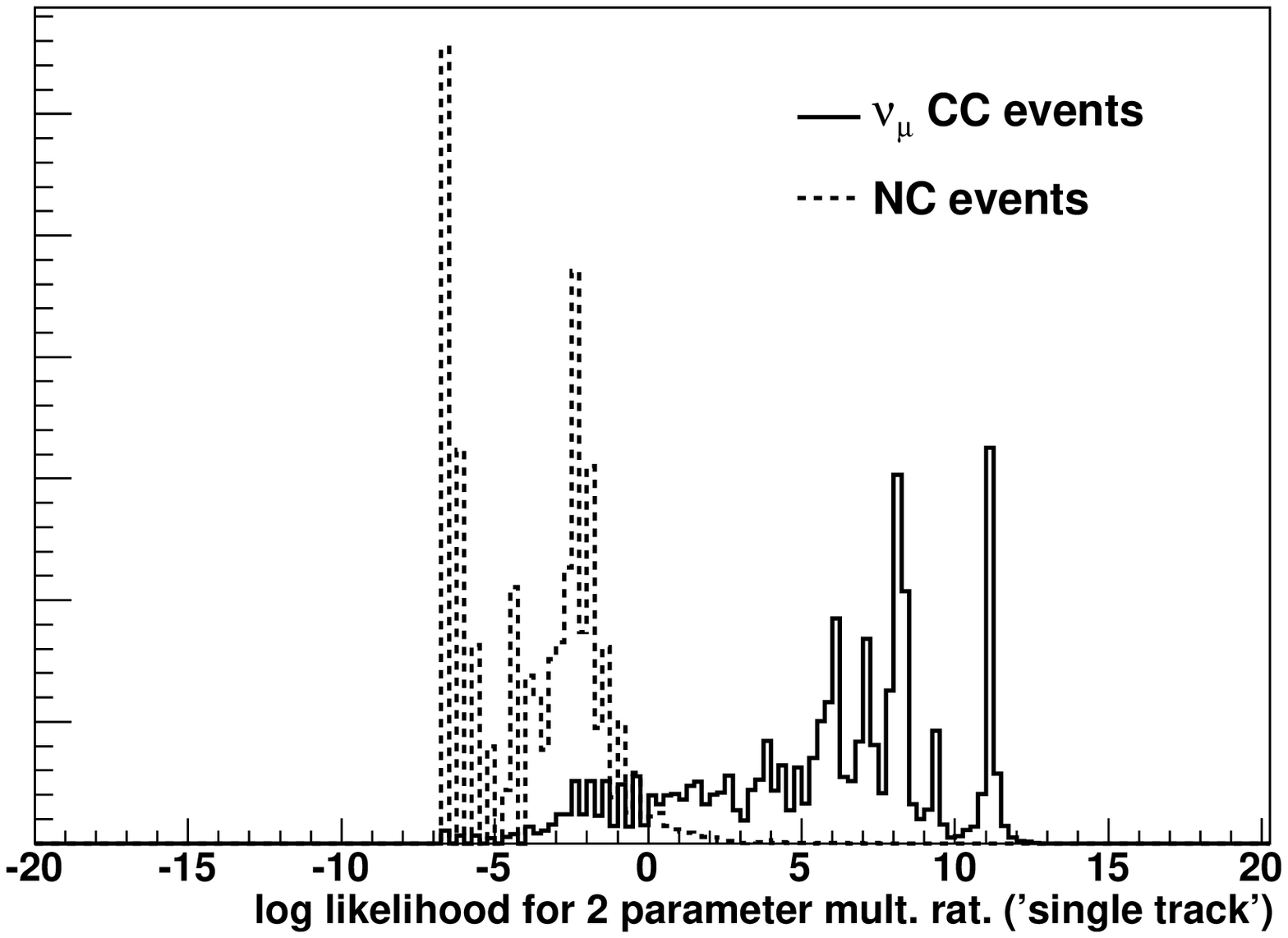}
  \end{center}
  \caption{\emph{Log likelihood discriminator distributions for energy fraction \textless 0.999, i) $\mathcal{L}_1$, defined in  
Eq.~\ref{eq:loglike1}, ii) $\mathcal{L}_2$, defined in Eq.~\ref{eq:loglike2} and iii)  for energy fraction $\geq$ 0.999, $\mathcal{L}_3$, 
defined in Eq.~\ref{eq:loglike3}.} } 
  \label{fig:log1}
\end{figure}

The likelihood ratio for each of the three observables was computed and combined in three main 
log likelihood discriminators described in Eqs.~\ref{eq:loglike1} to~\ref{eq:loglike3}:
\begin{eqnarray}
  \label{eq:loglike1}
    \mathcal{L}_1 &= &\log \left( \frac{l_{hit}^{CC}\times l_{frac}^{CC}\times l_{var}^{CC}}{l_{hit}^{NC}\times l_{frac}^{NC}\times l_{var}^{NC}} \right)\\
    \label{eq:loglike2}
    \mathcal{L}_2 &= &\log \left( \frac{l_{2D}^{CC}\times l_{hit}^{CC}}{l_{2D}^{NC}\times l_{hit}^{NC}} \right)\\
    \label{eq:loglike3}
    \mathcal{L}_3 &= &\log \left( \frac{l^{CC^{frac=1}}_{hit}\times l^{CC^{frac=1}}_{var}}{l^{NC^{frac=1}}_{hit}\times l^{NC^{frac=1}}_{var}} \right)
\end{eqnarray}
$\mathcal{L}_1$ is formed from the multiplication of the likelihoods mentioned above while $\mathcal{L}_2$ is formed by the 
multiplication of the $l_{hit}$ likelihood and a 2 dimensional likelihood of the variance and energy fraction ($l_{2D}=l_{hit}:l_{frac}$). 
$\mathcal{L}_1 \mbox{ or } \mathcal{L}_2$ are used when the energy fraction is less than 0.999 and $\mathcal{L}_3$ otherwise.
Distributions of these discriminators for samples of $\nu_\mu$ NC and CC events are shown in Fig.~\ref{fig:log1}.
 
%Considering the resultant values of these discriminators for groups of NC and CC events it can be seen (Fig.~\ref{fig:log1}) that 
%cutting here can separate the two classes of event efficiently. 
%The combination of more than one parameter in a single cut thus has the possibility 
%of reducing the backgrounds without significant loss of efficiency.

%--------------------------------------------------------------
\subsection{Fiducial cuts}
%--------------------------------------------------------------
\label{sec:fidcut}

Events originating near the edges of the detector can leave the sensitive volume. 
This will result not only in the loss of event energy and thus worsened energy resolution but, 
due to the shortening of the event, can cause a misidentification of the charge of a candidate. 
While the shortening of the event has the potential to reduce backgrounds from NC and $\nu_e$ CC as there should be less viable candidates, 
viable signal can also be lost, with a corresponding increase in charge misidentification background. 
Therefore, it is recomendable to apply a fiducial volume cut so that these pathologies are minimised. 
Specifically, events are rejected should their candidate have both its first hit within $50$~cm of the sides or back of 
the detector and its last within $10$~cm. In sec.~\ref{subSec::Inclusive}, the edge effects and their suppression are presented using $\overline{\nu}_\mu$ CC events as a model since they should affect little or no increase on NC and $\nu_e$ CC backgrounds and  any small variation should be of the same spectral form as those seen in $\overline{\nu}_\mu$ CC events.

%--------------------------------------------------------------
\subsection{Kinematic cuts}
%--------------------------------------------------------------
\label{PQT}
%considering the angular separation of the candidate momentum vector from that of the hadronic shower~\cite{CerveraVillanueva:2008zz}. 

\begin{figure}
  \begin{center}$
    \begin{array}{cc}
      \includegraphics[scale=0.35]{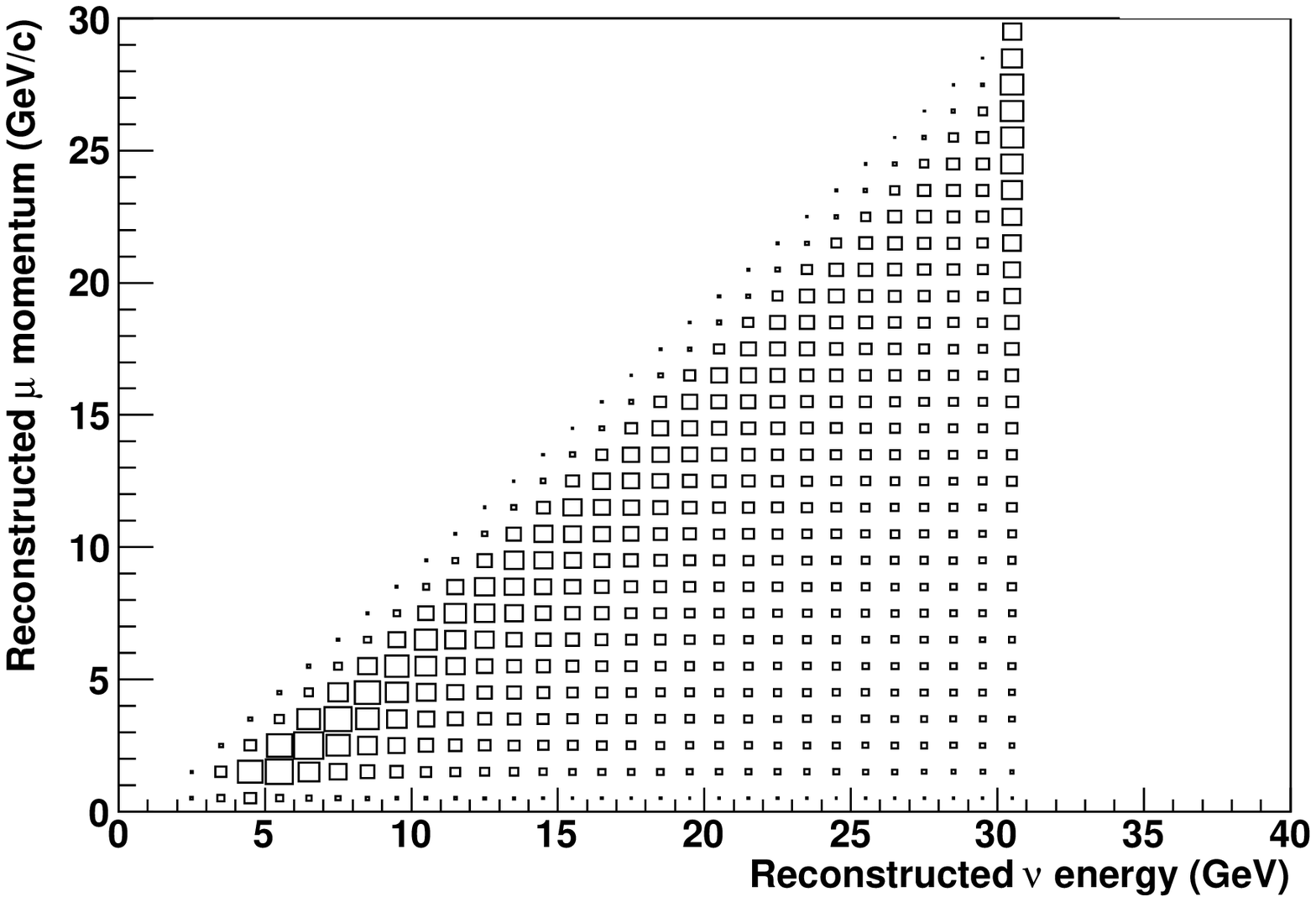} &
      \includegraphics[scale=0.35]{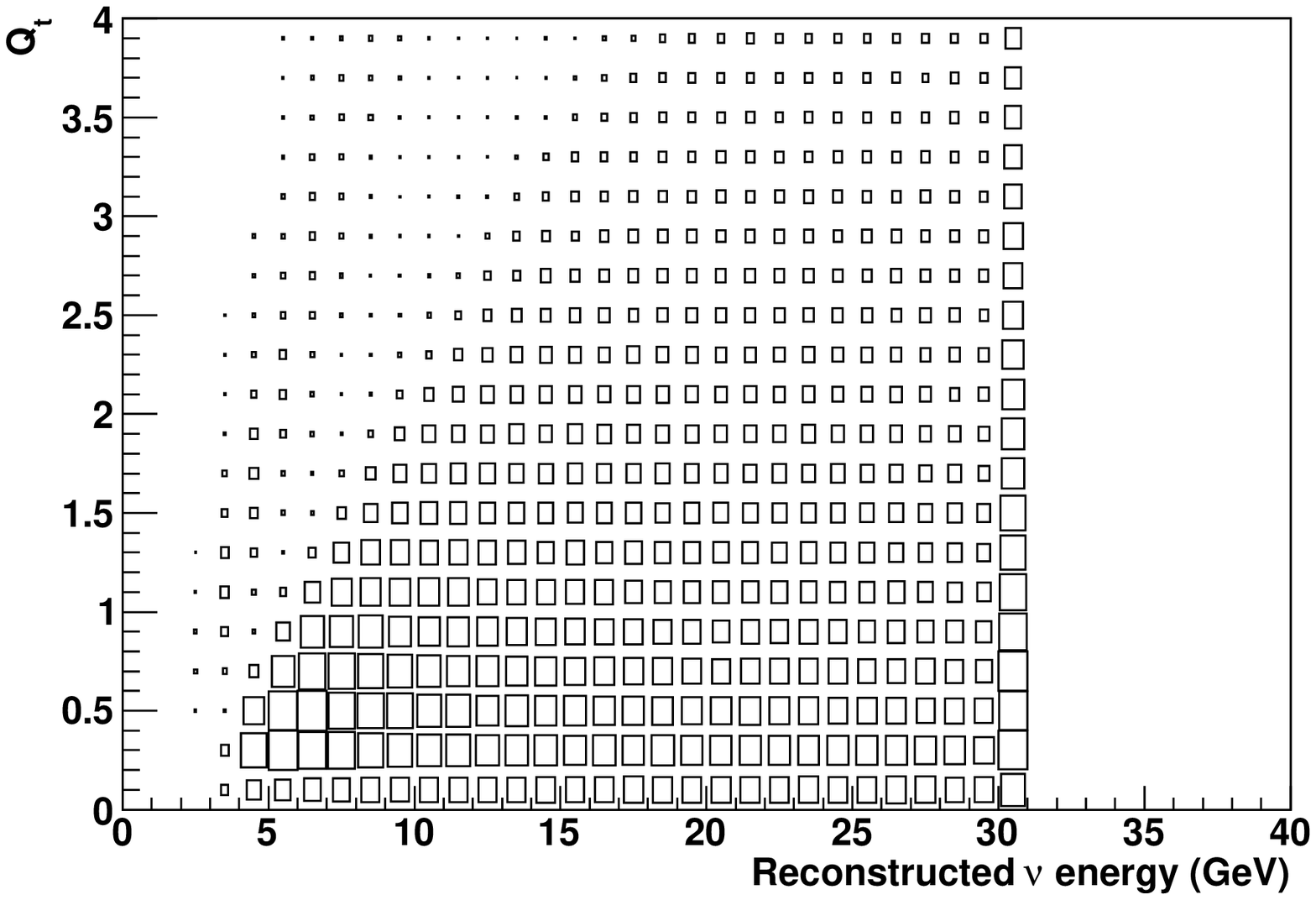} \\
      \includegraphics[scale=0.35]{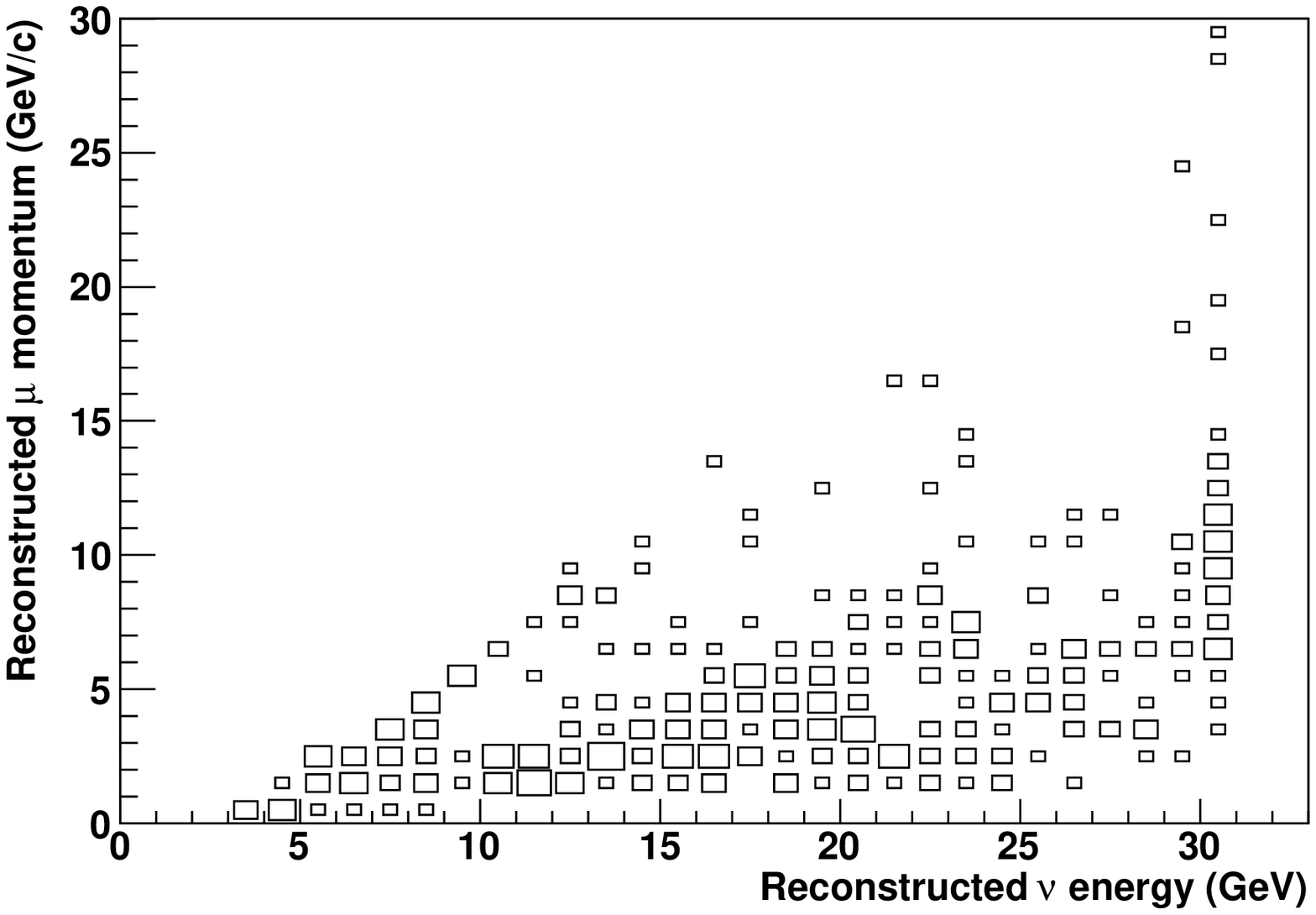} &
      \includegraphics[scale=0.35]{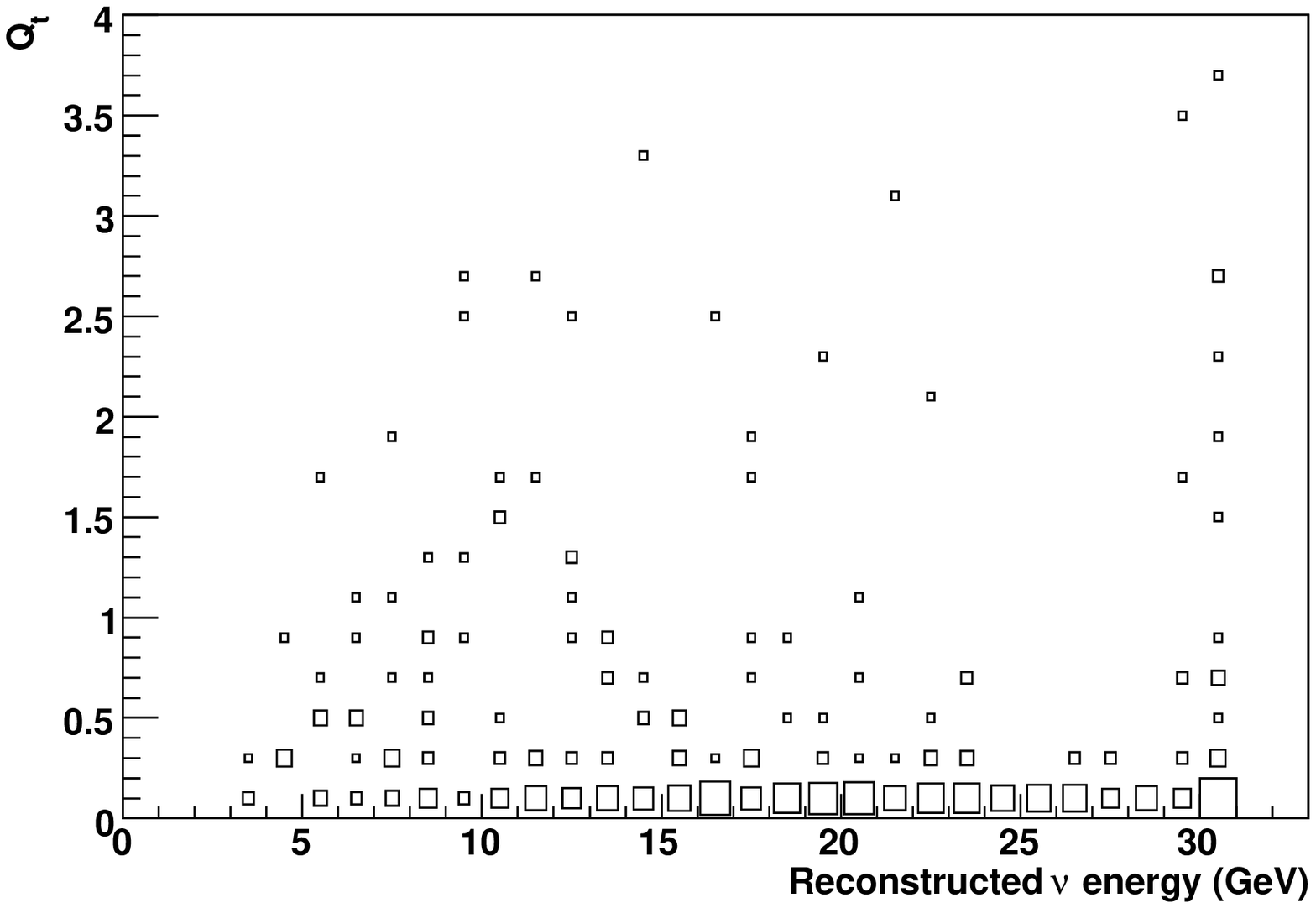} \\
      \includegraphics[scale=0.35]{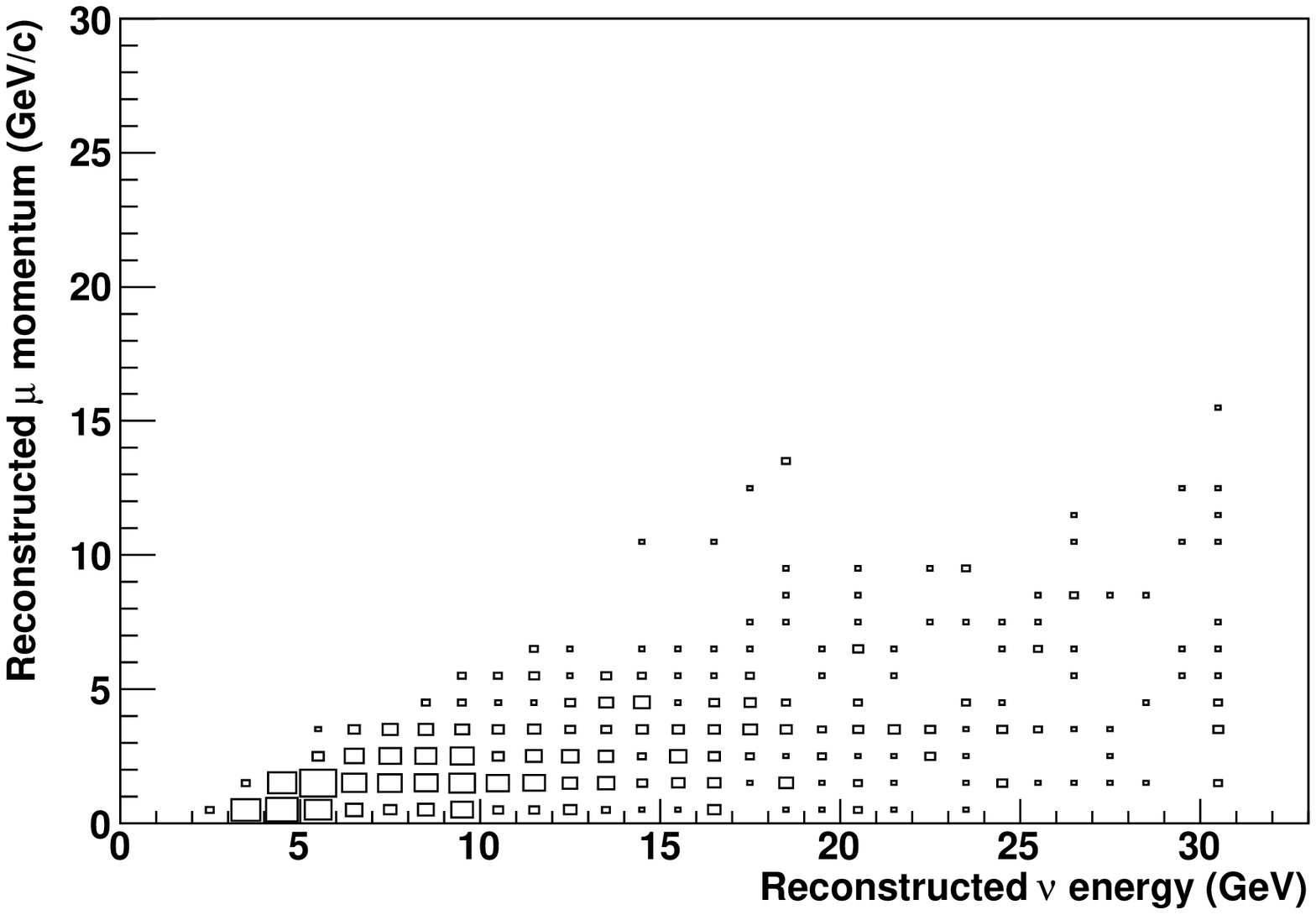} &
      \includegraphics[scale=0.35]{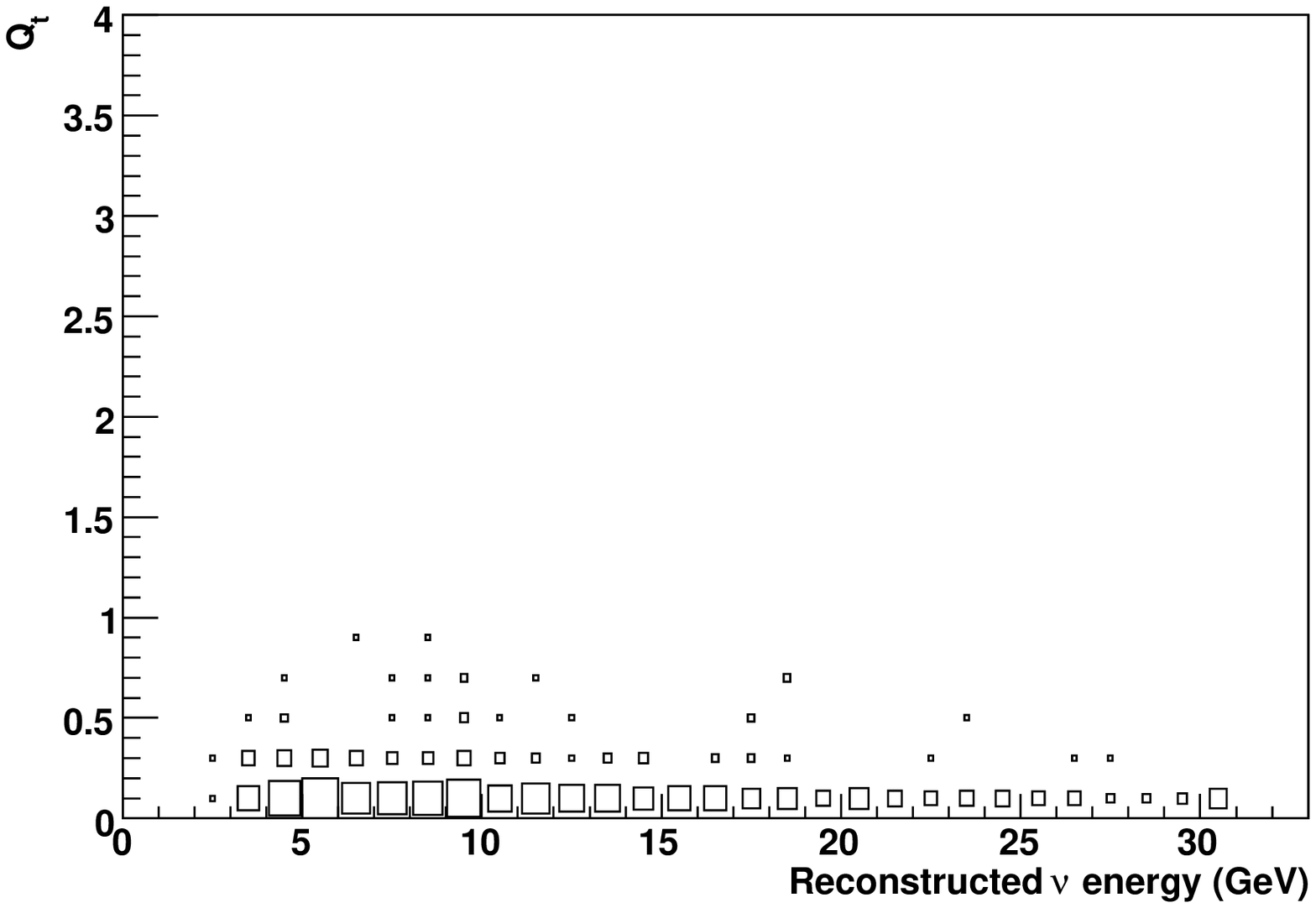} \\
      \includegraphics[scale=0.35]{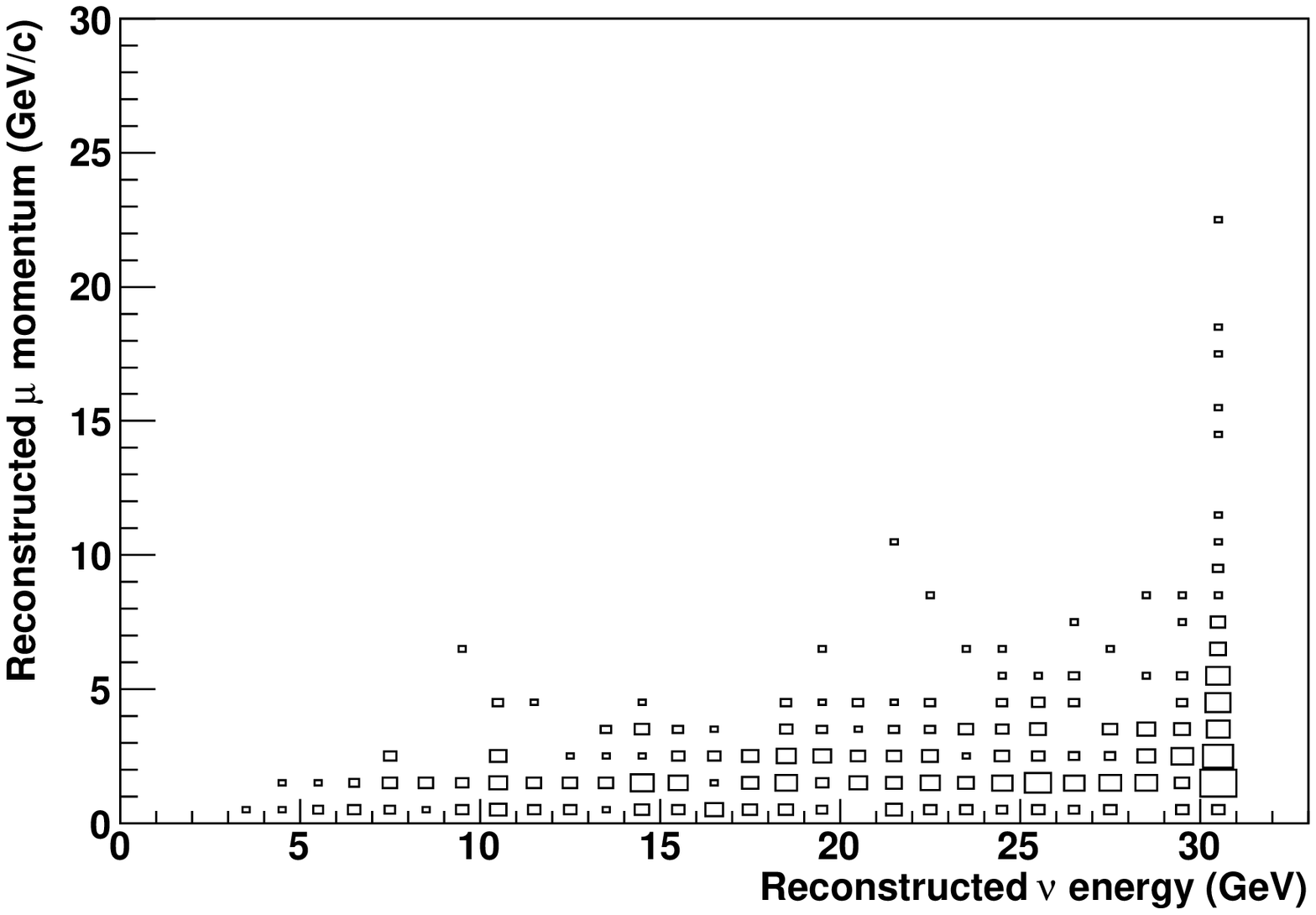} &
      \includegraphics[scale=0.35]{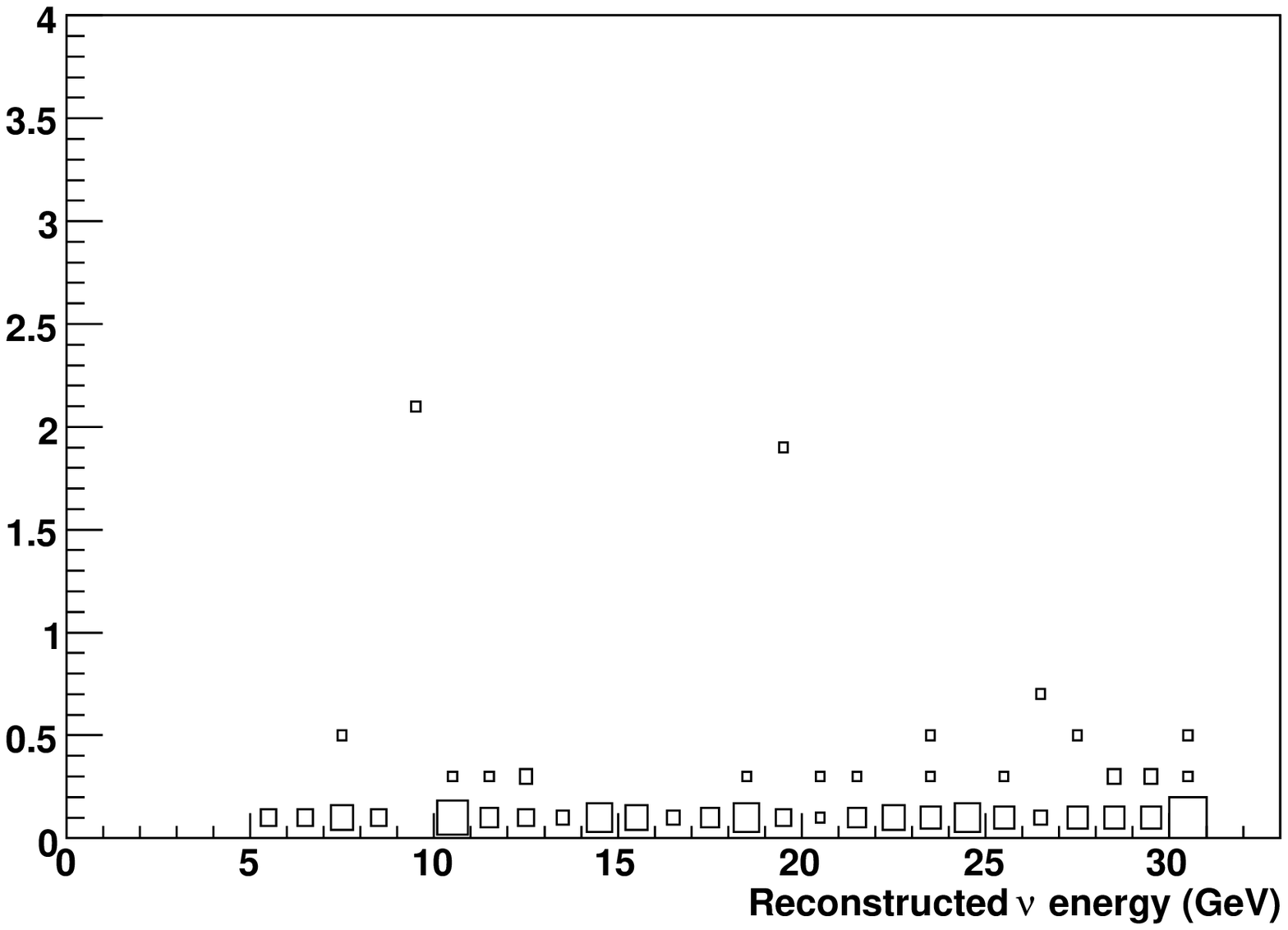} \\
    \end{array}$
  \end{center}
  \caption{\emph{Distributions of kinematic variables: Reconstructed muon momentum (left) and $Q_t$ variable in GeV/c units (right) versus reconstructed neutrino energy, 
for (top$\rightarrow$bottom) signal, $overline{\nu}_\mu$ CC backgrounds, NC and $\nu_e$ CC.}} %Anselmo. Where is the numucc background from hadron decays ?
  \label{fig:PQT}
\end{figure}

Considering the remaining signal and background after applying all cuts described above, the $Q_t$ (see Sec.~\ref{HadFit}) 
and muon candidate momentum ($P_\mu$) distributions are those shown in Fig.~\ref{fig:PQT}. A clear separation between signal 
and background events is observed. In particular, background events are concentrated at very low $Q_t$, while the signal exhibits much larger $Q_t$ 
values. In order not to reduce the efficiency at low neutrino energy, cuts on these two variables are only applyed for reconstructed 
neutrino energy ($E_\nu$) above $7$~GeV. The applyed cuts are those of Eq.~\ref{eq:kinCut}: 

\begin{equation}
  \label{eq:kinCut}
  P_\mu \geqslant 0.2 \cdot E_\nu \,\,\, \mbox{ and } \,\,\, Q_t > 0.25~\mbox{GeV/c} \,\,\,\,\,\, \mbox{ for } E_\nu> 7~\mbox{GeV}.
\end{equation}

%--------------------------------------------------------------
\subsection{Summary of analysis cuts}
%--------------------------------------------------------------

As will be dicussed in the next section the most succesful set of cuts is given below in table~\ref{tab:ana_cuts}.

\begin{table}[ht]
  \begin{center}
    \begin{tabular}{l|c}
       Cut type & Cut value \\
      \hline
      \hline
      Fiducial               &  if $|r^{first}_i-r^{det}_i|\leqslant 50~cm$, $|r^{last}_i-r^{det}_i|>10~cm$ \\  
                             &    for $r_i=x,y,z$ \\ 
      \hline
      Track quality          & $\mathcal{L}_{q/p} > 2.0$ and $\chi^2_{prob} > 0.9999$  \\
      \hline
      $\nu_\mu$ CC selection & $\mathcal{L}_1 > 0$ for  $l_{frac} <0.999$ \\
                             & $\mathcal{L}_3 > 0$ for  $l_{frac} \geq 0.999$ \\
      \hline
      Kinematic              & $P_\mu \geqslant 0.2 \cdot E_\nu$  and $Q_t > 0.25$ for $E_\nu> 7~GeV$ \\
    \end{tabular}
  \end{center}
  \caption{\emph{Summary of analysis cuts.}}
  \label{tab:ana_cuts}
\end{table}

%**********************************************************************
\section{Analysis Results}
%**********************************************************************
\label{res}
Using a large data set and the analyses described above the efficiency and rejection power of MIND has been studied. 

%--------------------------------------------------------------
\subsection{$\overline{\nu}_\mu$ charge current interactions}
%--------------------------------------------------------------

The background from $\overline{\nu}_\mu$ interactions can be separated in two different contributions: 
i)  fake wrong-sign muons from charge misidentification of the primary muon (mainly) and from pion to muon confusion, and   
ii) true wrong-sign muons from the decay of hadrons.

%--------------------------------------------------------------
\subsubsection{Incorrect charge assignment}
%--------------------------------------------------------------
\label{misID}
The charge misidentification background was studied using $\overline{\nu}_\mu$ interactions where events containing 
hadronic decays to $\mu^-$ were excluded to be considered separately (Sec.~\ref{wsDec}).
An event is considered background if a candidate is successfully extracted and fitted with charge opposite 
to that of the true primary muon. Background events are mainly due to incorrect charge assignment to the true primary 
muon (due to multiple scattering or impurity of the candidate), but have a small contribution from penetrating hadrons (mainly pions) 
which are identified as muon candidates when the true primary muon has low momentum and is not correctly identified. 

%The purity and nature and topographic form of the candidates in the remaining background events 
%will be dicussed along with possible further reductions in this background in Sec.~\ref{sum}.

\begin{figure}
  \begin{center}$
    \begin{array}{cc}
      \includegraphics[scale=0.37]{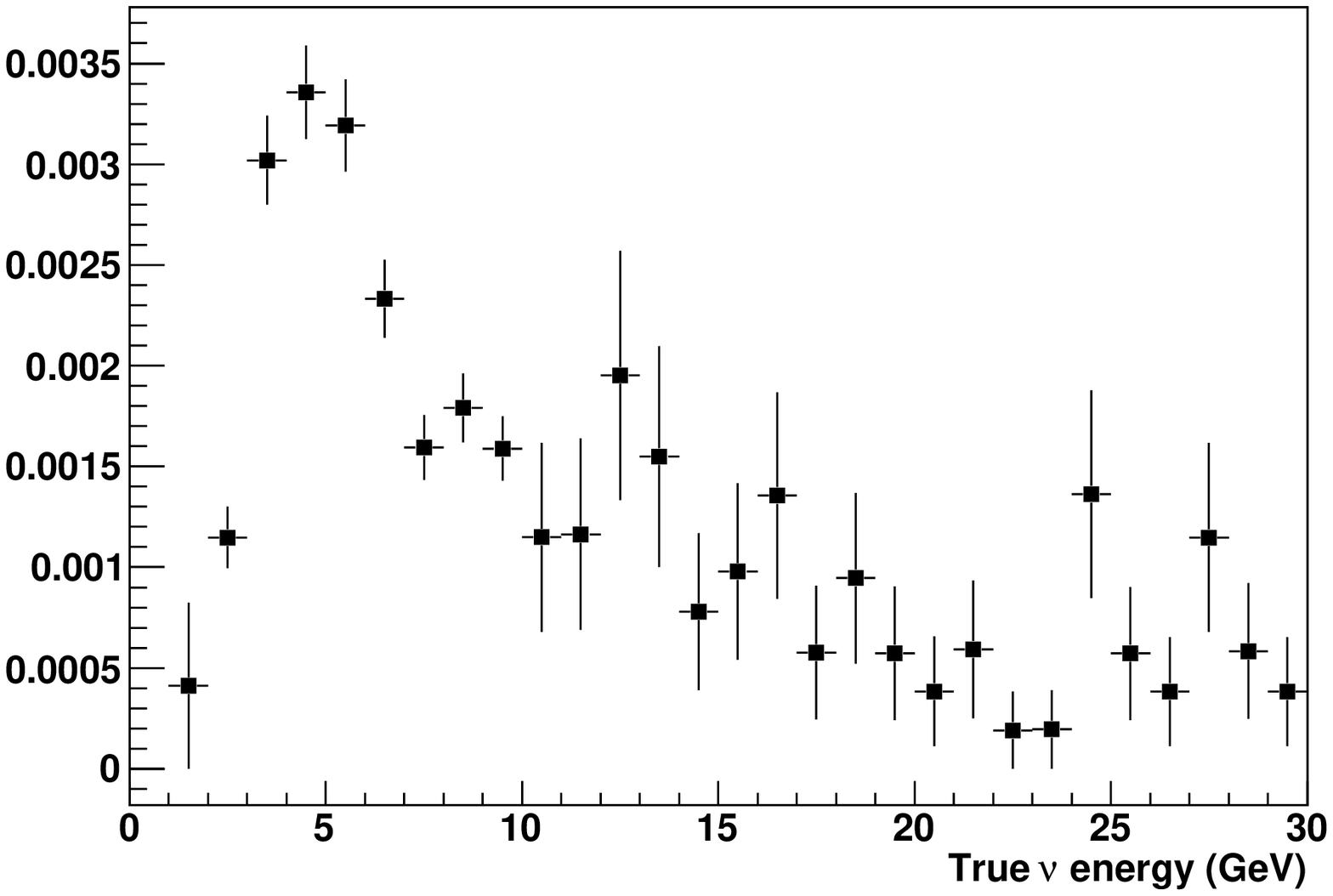} &
      \includegraphics[scale=0.37]{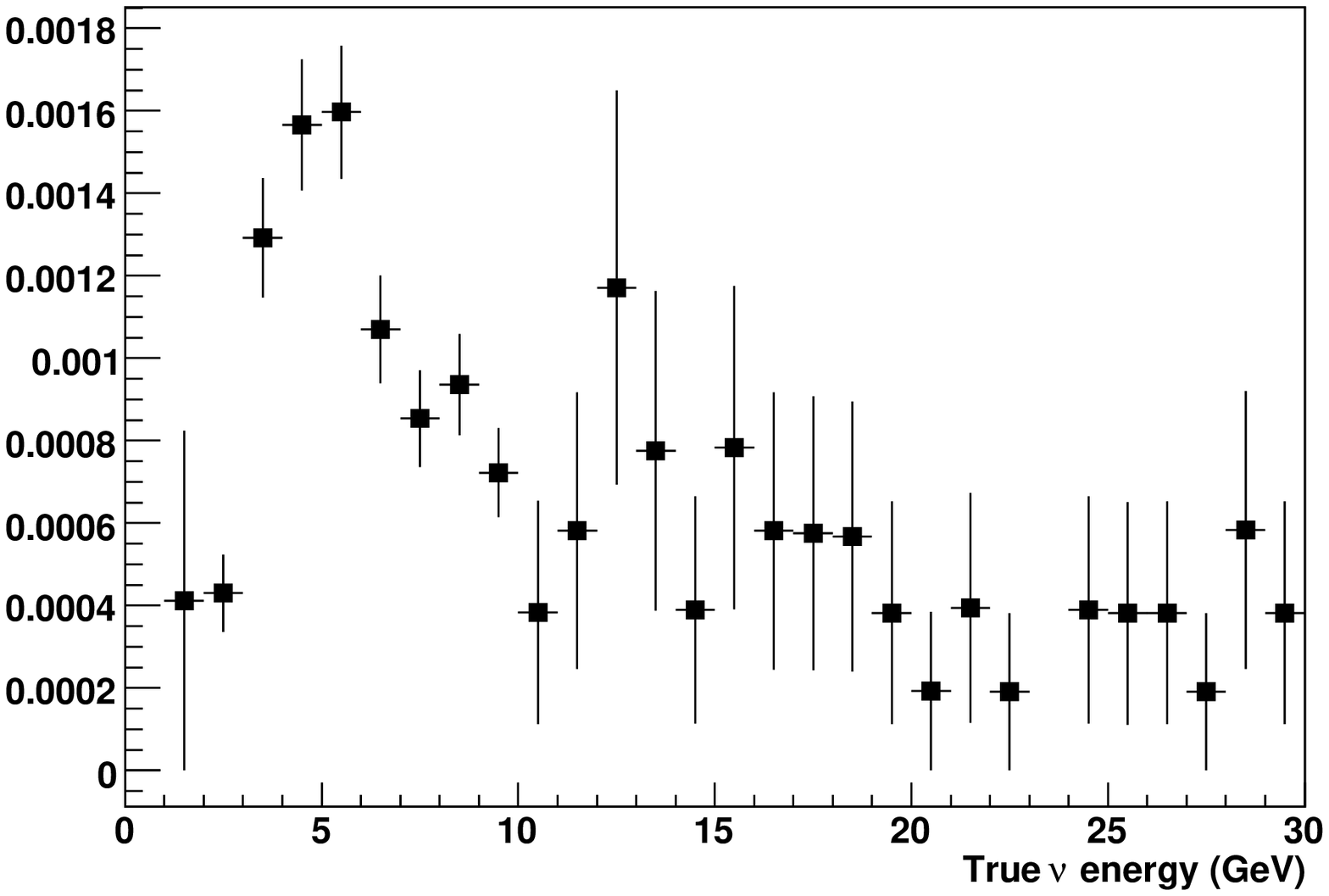}
    \end{array}$
  \end{center}
  \caption{\emph{Charge mis-assignment and hadron to muon mis-identification background i) with track quality cuts only and ii) including $\nu_\mu$ CC selection cuts.}}
  \label{fig:numu_charge_bkg}
\end{figure}

As shown in Fig.~\ref{fig:numu_charge_bkg}-(left) this background can be efficiently suppressed by cutting on the track quality variables, 
described in Sec.~\ref{qualcut}. Further rejection is obtained by applying $\nu_\mu$ 
CC selection cuts (see Fig.~\ref{fig:numu_charge_bkg}-(right)).
% since in this case the contribution from penetrating pions in CC interactions is significantly suppresed. (???)

%--------------------------------------------------------------
\subsubsection{Wrong sign muons from hadron decays}
%--------------------------------------------------------------
\label{wsDec}
The production and decay of negatively charged mesons in the hadronic part of a DIS interaction 
has high probability to produce a $\mu^-$. Particulary mesons containing charm will decay promptly and 
produce high energy muons which can be selected as primary muon candidates when the true primary muon is not correctly identified 
(in general when it has low momentum). Suppression of this background is particulary important due to the 
high level of uncertainty on the value of the charm production cross section~\cite{:2008xp}. Track quality and $\nu_\mu$ CC selection cuts are effective in reducing 
this type of background for low neutrino energies, but the suppresion of high energy background requires further cuts, which are described below.   
%The current analysis is capable of suppressing this background to a similar level as others described above, Fig.~\ref{fig:wslike}. 

%These events constitute a significant background in their own right and thus merit separate consideration.

\begin{figure}
  \begin{center}$
    \begin{array}{cc}
      \includegraphics[scale=0.37]{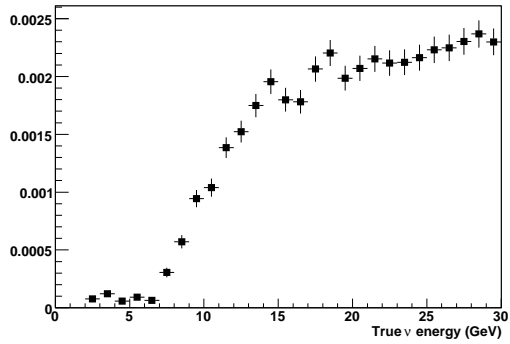} 
    \end{array}$
  \end{center}
  \caption{\emph{Wrong sign muon decay background with track quality and $\nu_\mu$ CC selection cuts.}}
  \label{fig:numu_hadron_bkg}
\end{figure}

%--------------------------------------------------------------
\subsubsection{Inclusive $\overline{\nu}_\mu$ background with fiducial and kinematic cuts}
\label{subSec::Inclusive}
%--------------------------------------------------------------

\begin{figure}[ht]
  \begin{center}$
    \begin{array}{cc}
      \includegraphics[scale=0.35]{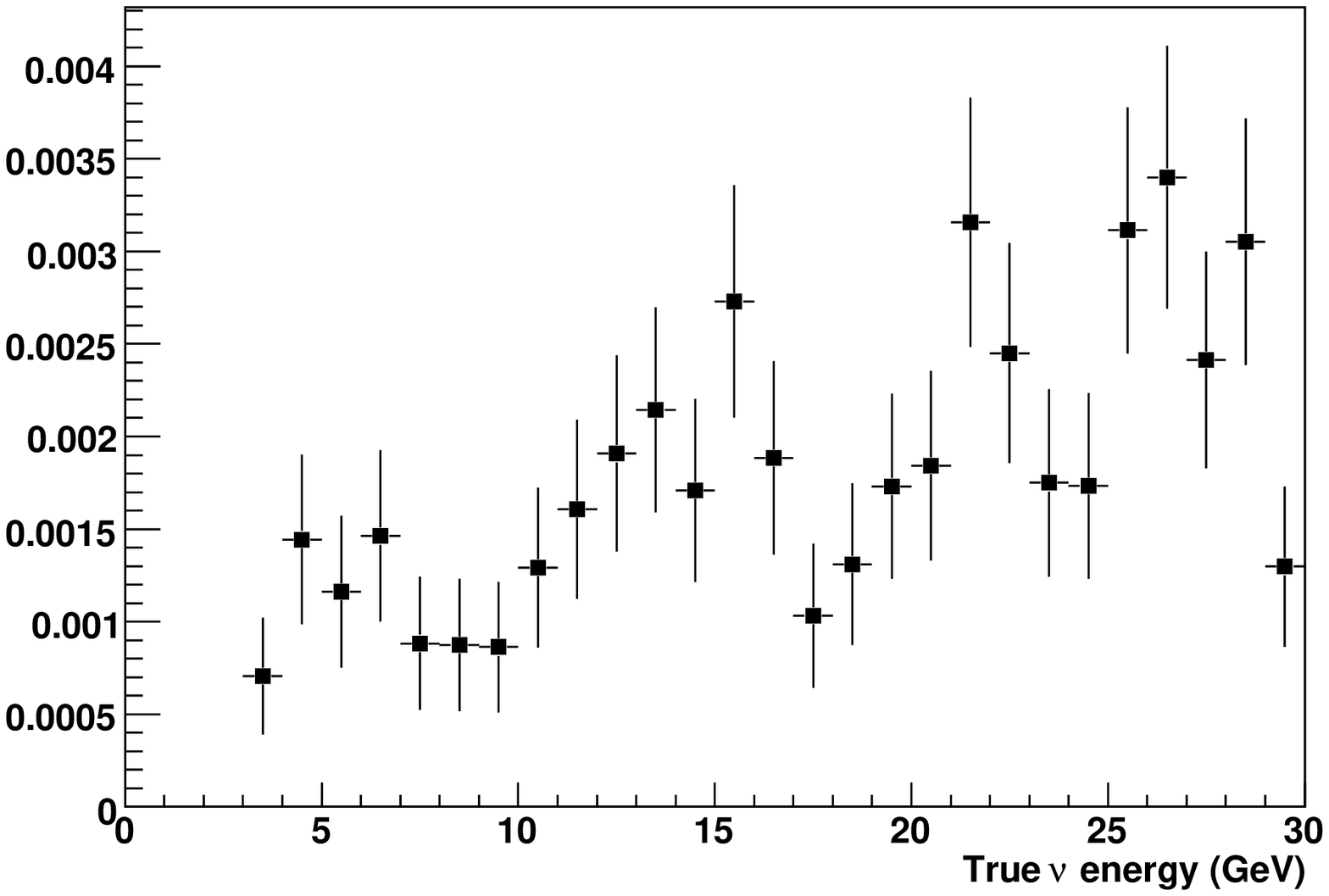} &
      \includegraphics[scale=0.35]{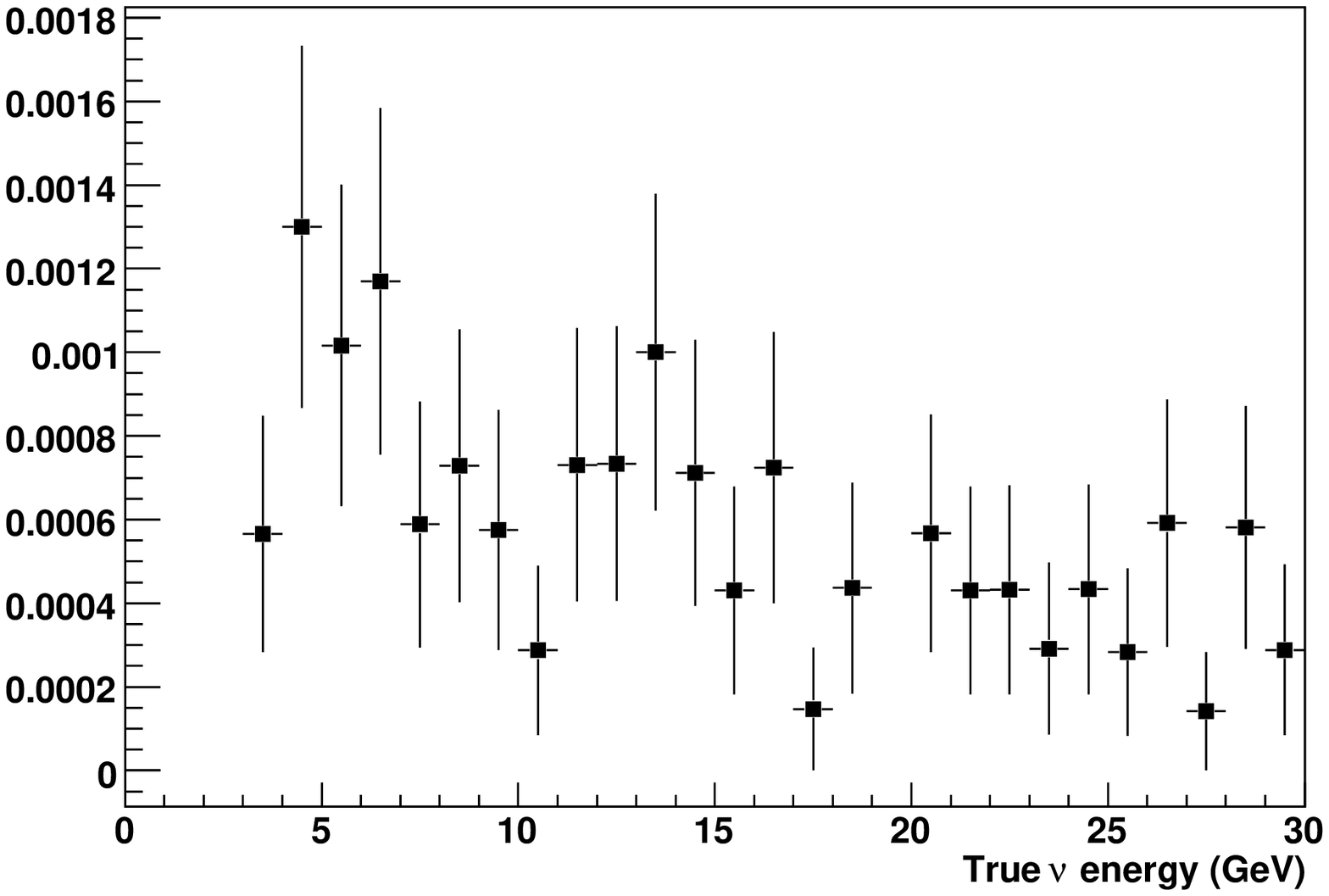} \\
    \end{array}$
  \end{center}
  \caption{\emph{Expected background from $\overline{\nu}_\mu$ CC interactions when events 
are randomly generated in the entire detector: i) after track quality, $\nu_\mu$ CC selection and fiducial cuts, and ii) including kinematic cuts.}}
  \label{fig:numu_bkg}
\end{figure}

Considering a set of  $\overline{\nu}_\mu$ CC events generated randomly throughout the detector volume 
the inclusive background from this type of interaction has been studied. An additional cut on those 
events with a candidate failing the fiducial volume cut, 
defined in Sec.~\ref{sec:fidcut}, is used to suppress background 
caused by edge effects. As can be seen in Fig.~\ref{fig:numu_bkg}-(left), the additional background 
introduced by edge effects is almost compensated by the fiducial cut, leading to a inclusive $\overline{\nu}_\mu$ 
background similar to the addition of the ones shown in Figs.~\ref{fig:numu_charge_bkg}-(right) and \ref{fig:numu_hadron_bkg}. 

The high energy background of Fig.~\ref{fig:numu_bkg}-(left) is mainly due to very hard muons from the decay of charm mesons. Fortunately these muons, due to their decay origin, tend to be imbedded in the hadron shower unlike a true primary muon. Thus the $Q_t$ variable should be very effective in rejecting this kind of event. As can be seen in Fig.~\ref{fig:numu_bkg}-(right) the kinematic cuts afford a sizeable suppression, particularly at higher neutrino energy. 

%--------------------------------------------------------------
\subsection{Neutral current interactions}
\begin{figure}
  \begin{center}$
    \begin{array}{cc}
      \includegraphics[scale=0.35]{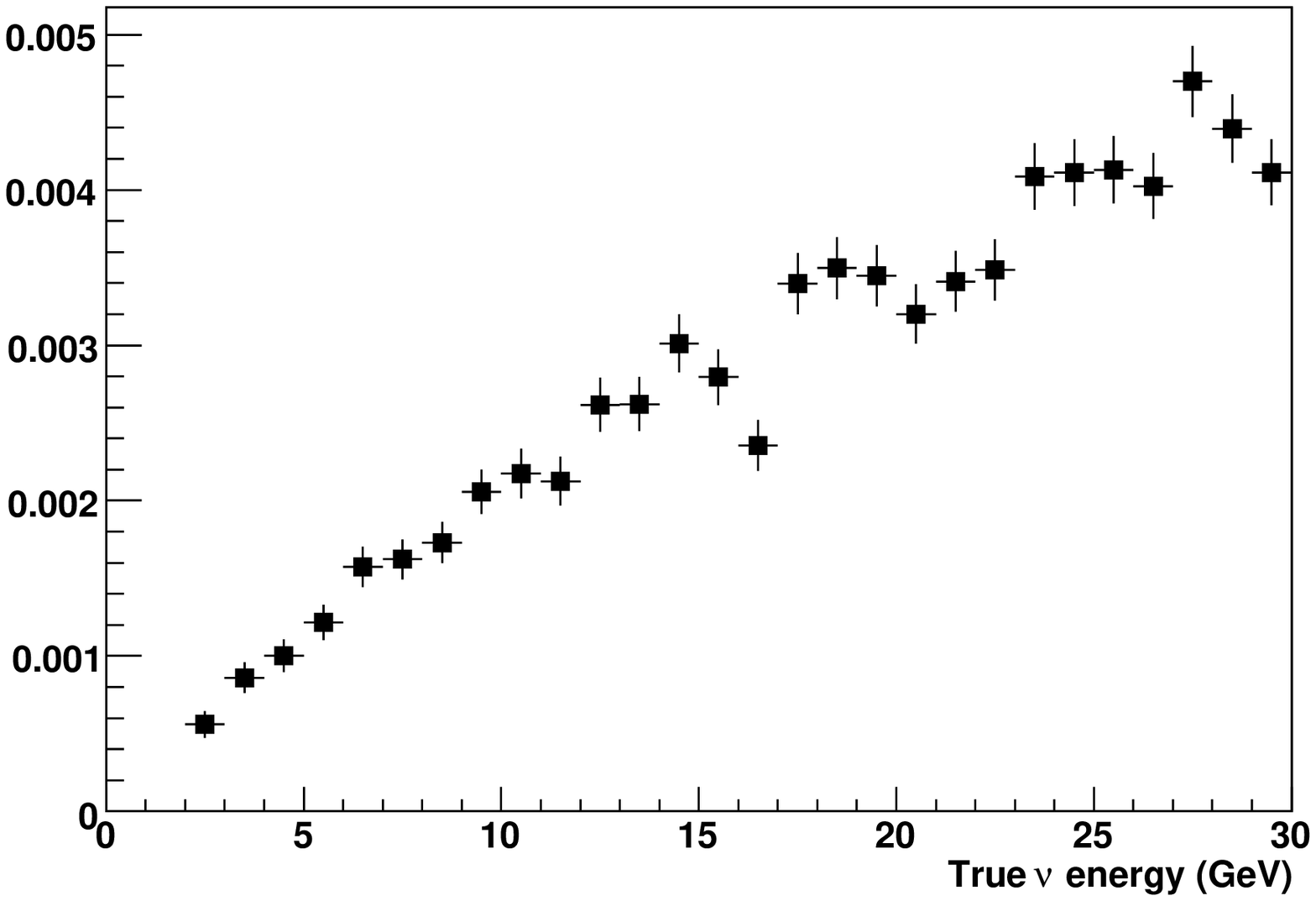}  &
      \includegraphics[scale=0.35]{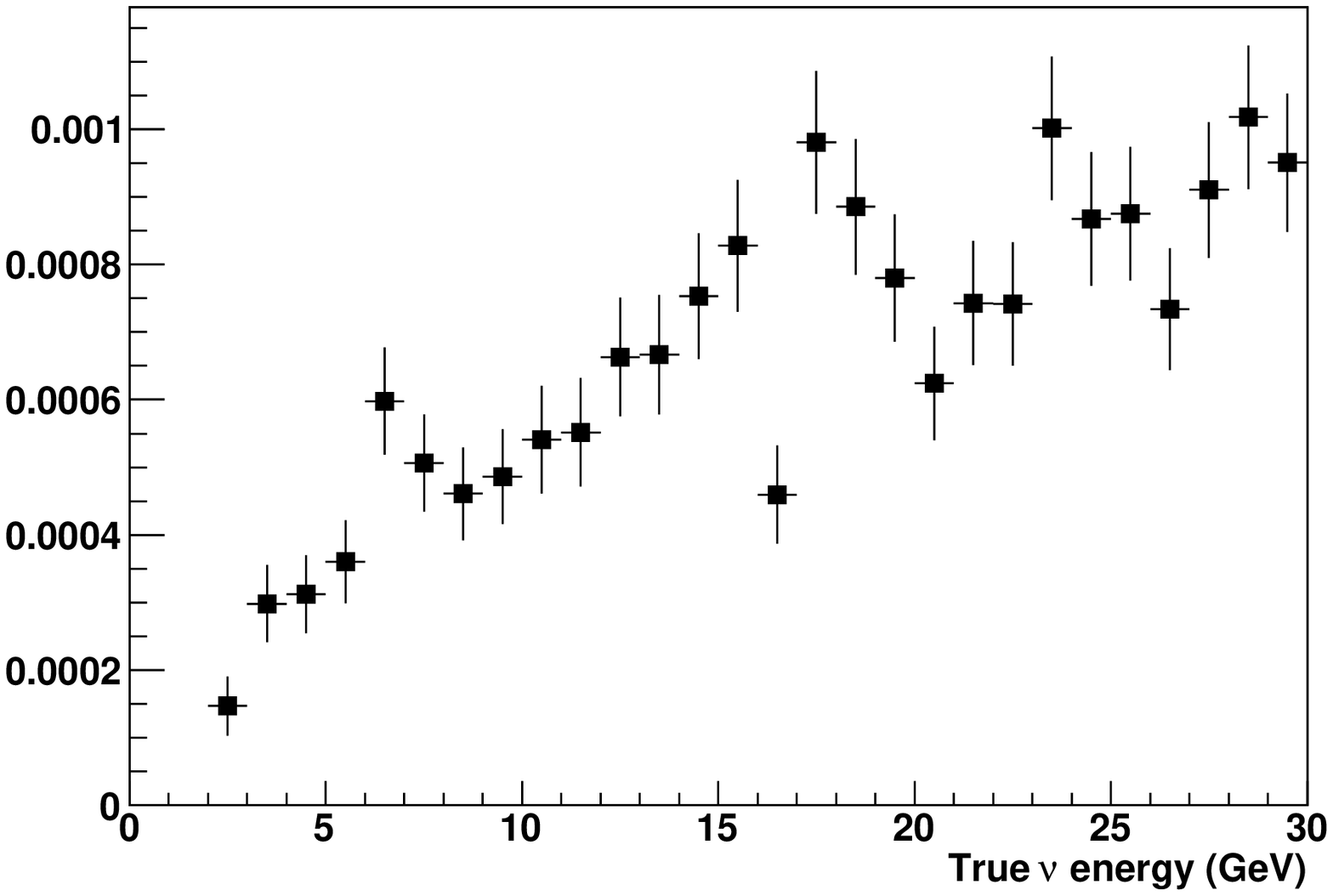} \\
      \includegraphics[scale=0.35]{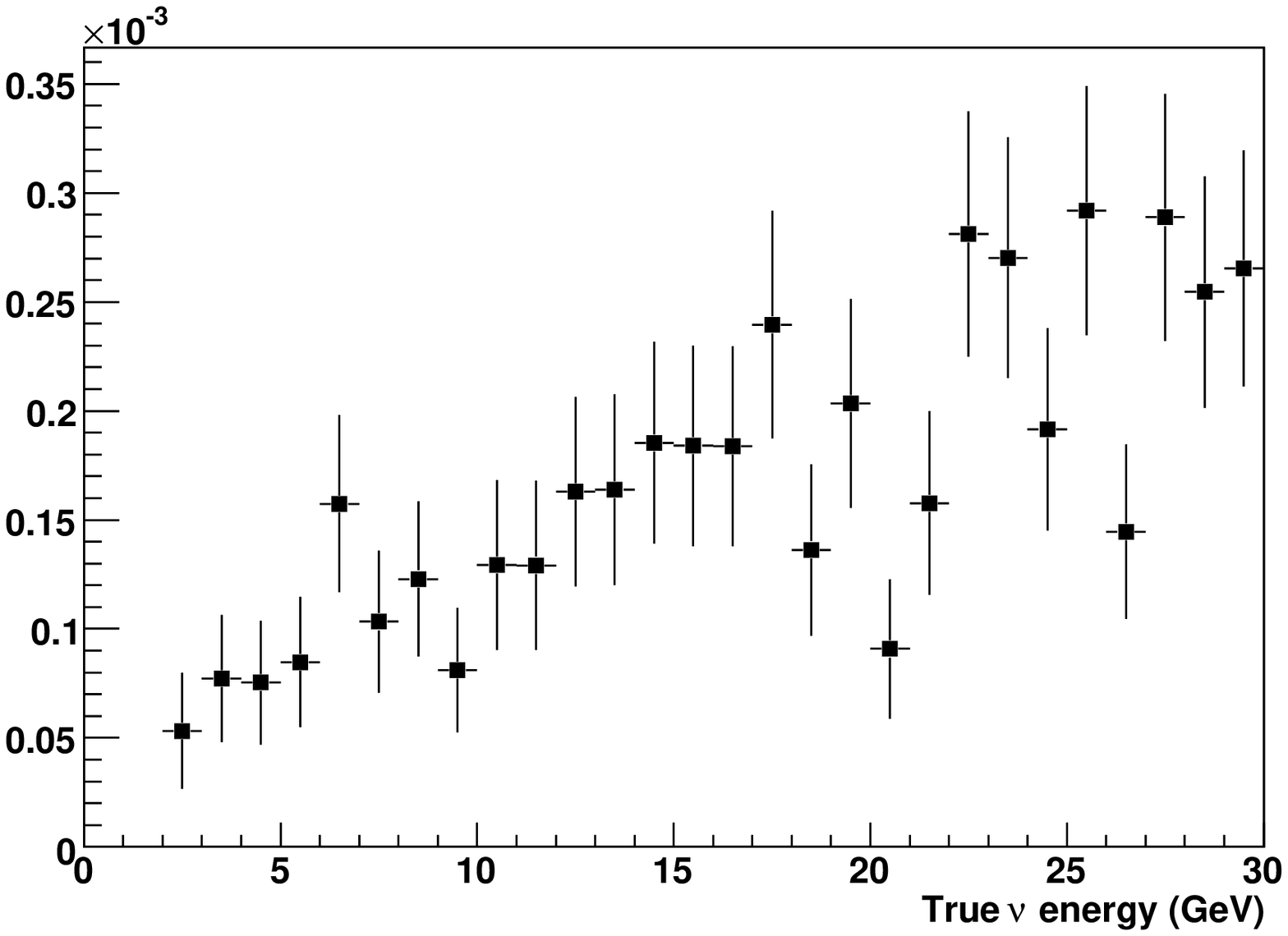} &
      \includegraphics[scale=0.35]{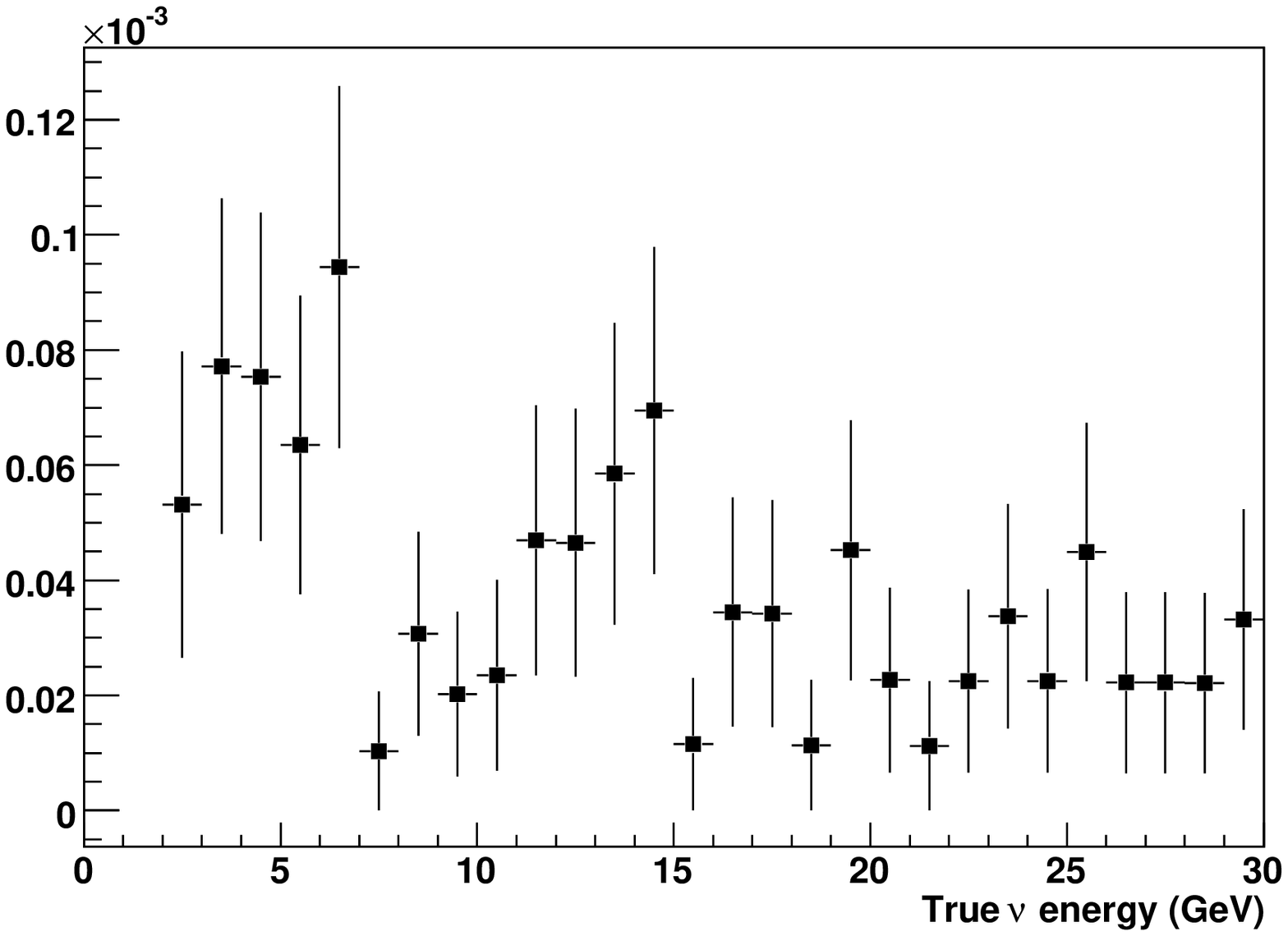}
    \end{array}$
  \end{center}
  \caption{\emph{Expected background from $\overline{\nu}_\mu$ NC interactions, i) with track quality cuts (Eq.~\ref{eq:qualcut1}) only, 
ii) including $\nu_\mu$ CC selection cuts, iii) substituting track quality cuts of Eq.~\ref{eq:qualcut1} by those of 
Eq.~\ref{eq:qualcut2} and iv) including kinematic cuts.}}
  \label{fig:NC_bkg}
\end{figure}

%--------------------------------------------------------------
\label{NCback}
Neutral current interactions should be of the same nature for all species. Background events will 
tend to originate from penetrating pions or muons from the decay of hadrons. Moreover, since there will always be 
missing energy in the event, those events successfully fitted will tend to be reconstructed at 
lower energy than the true neutrino energy. As such and due to the large amount of NC events expected 
in the detector, this background must be suppressed efficiently. Fig.~\ref{fig:NC_bkg} shows the evolucion of the NC background 
when different cuts are included.

%Application of the quality cuts alone cannot reduce the 
%NC background sufficiently, as shown in Fig.~\ref{fig:NC}.

%Performing, in addition to the track quality cuts, the $\nu_\mu$ CC selection cuts (as described in~\ref{logLike}), 
%allows for further background suppression (see Fig.~\ref{fig:NC}). Alternatively, only allowing those trajectories 
%with a calculated $\mathcal{L}_{q/p}$ greater than $1.0$ and a $\mathcal{L}_1 \mbox{\textbackslash} \mathcal{L}_3$ value greater than $0$ it was found that 
%further reduction could be achieved (Fig.~\ref{fig:NC}). Further background rejection is obtained when appliying kinematic cuts.

%--------------------------------------------------------------
\subsection{$\nu_e$ charge current interactions}
%--------------------------------------------------------------
\label{eback}
The interactions of $\nu_e$ present in the beam can also produce some background to the signal. 
While the electron itself will be stopped quickly and will shower far more than a muon, penetrating 
pions or decay muons originating in the hadronic shower can be mistaken for primary muons.

\begin{figure}
  \begin{center}$
    \begin{array}{cc}
      \includegraphics[scale=0.37]{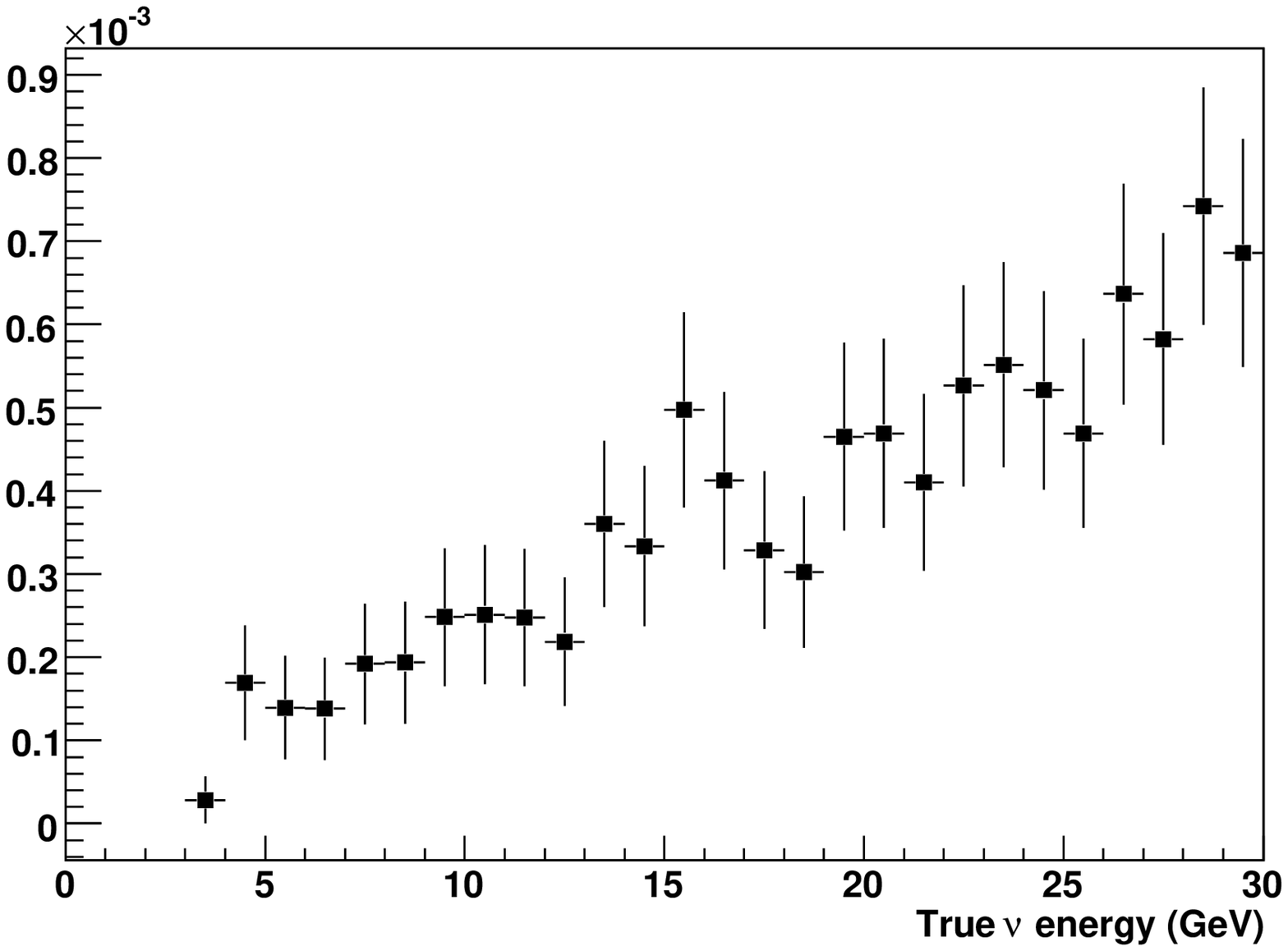} &
      \includegraphics[scale=0.37]{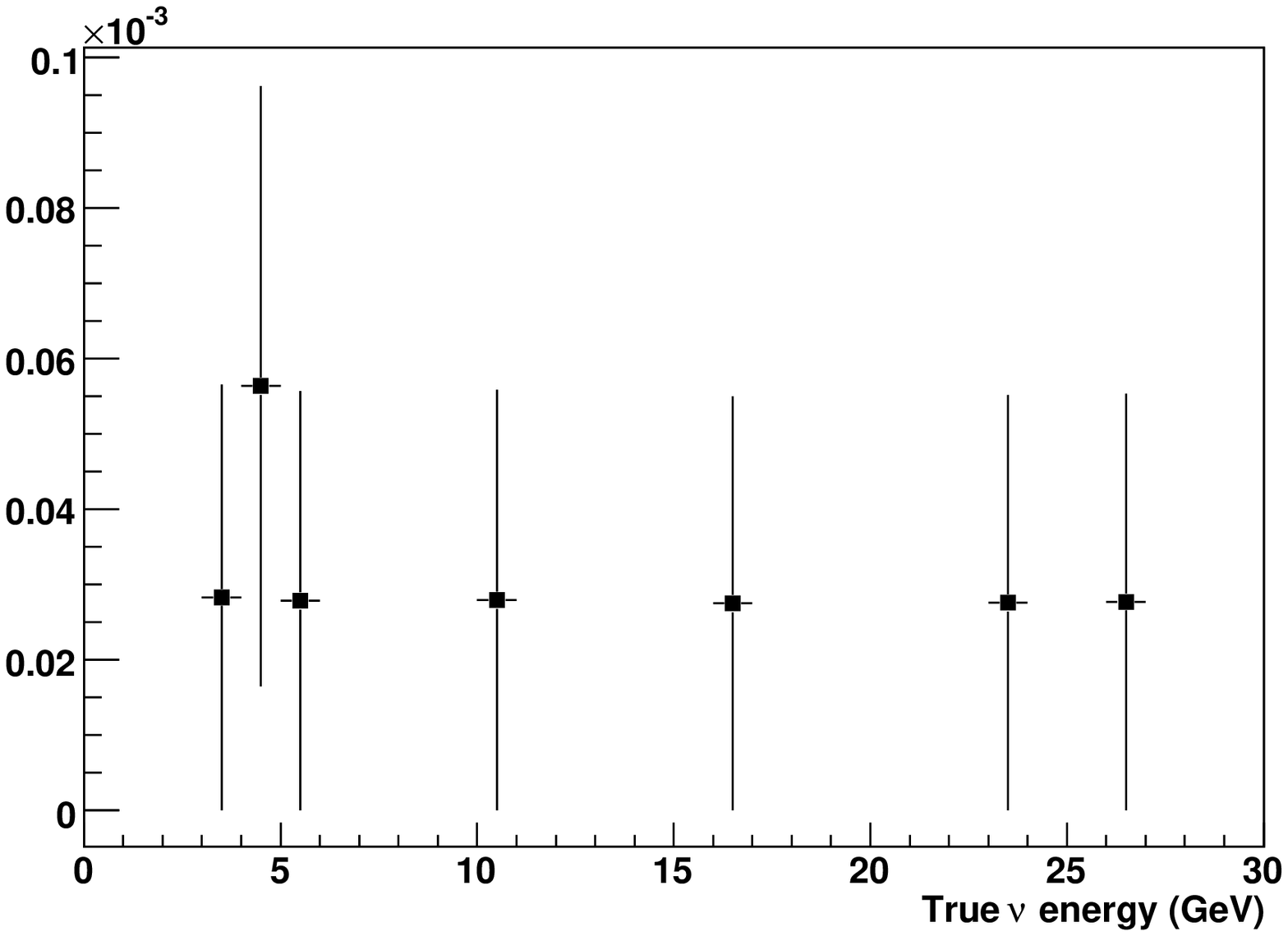} 
    \end{array}$
  \end{center}
  \caption{\emph{$\nu_e$ CC background to golden channel signal, i) with track quality and $\nu_\mu$ CC selection cuts, and ii) including kinematic cuts.}}
  \label{fig:nue_bkg}
\end{figure}

After application of track quality and $\nu_\mu$ CC selection cuts it can be seen in Fig.~\ref{fig:nue_bkg}-(left) that the $\nu_e$ CC background 
can be reduced to a similar level as the NC background when the same cuts are applied. Moreover, excellent rejection for high energy neutrinos is obtained by applying kinematic cuts, as shown in Fig.~\ref{fig:nue_bkg}-(right). %Anselmo
This is because the candidate muon in $\nu_e$ events tends to have lower momentum than in NC events, as shown in Fig.~\ref{fig:PQT} .

%However, the 
%energy reconstruction of the faked $\nu_\mu$ results in less low energy pollution due to the 
%lack of missing energy in the form of the primary neutrino. In real terms this will, however, 
%constitute a greater level of absolute background due to the larger cross section for $\nu_e$ 
%CC interactions. However, and excellent rejection for high energy neutrinos is obtained 
%by appliying kinematic cuts, as shown in ...

%--------------------------------------------------------------
\subsection{Summary}
%--------------------------------------------------------------
\label{sum}

\begin{table}[ht]
  \begin{center}
    \begin{tabular}{|c|c|c|c|c|}
      \hline
      & Total & $0-5$~GeV & $5-10$~GeV & $10-30$~GeV \\
      \hline
      \hline
      $\overline{\nu}_\mu$ CC & $5.5\times10^{-4}$ & $6.6\times10^{-4}$ & $8.2\times10^{-4}$ & $4.6\times10^{-4}$\\
      $\nu_e$ CC & $7.8\times10^{-6}$ & $2.6\times10^{-5}$ & $5.5\times10^{-6}$ & $5.5\times10^{-6}$\\
      $\overline{\nu}_\mu + \nu_e$ NC & $3.8\times10^{-5}$ & $6.8\times10^{-5}$ & $4.3\times10^{-5}$ & $3.1\times10^{-5}$ \\
      $\nu_\mu$ (signal) & $0.64$ & $0.25$ & $0.66$ & $0.69$\\
      \hline
    \end{tabular}
  \end{center}
  \caption{\emph{Summary of expected fractional signal and background with true neutrino energy.}}
  \label{tab:backsum}
\end{table}

\begin{figure}
  \begin{center}$
    \begin{array}{cc}
      \includegraphics[scale=0.35]{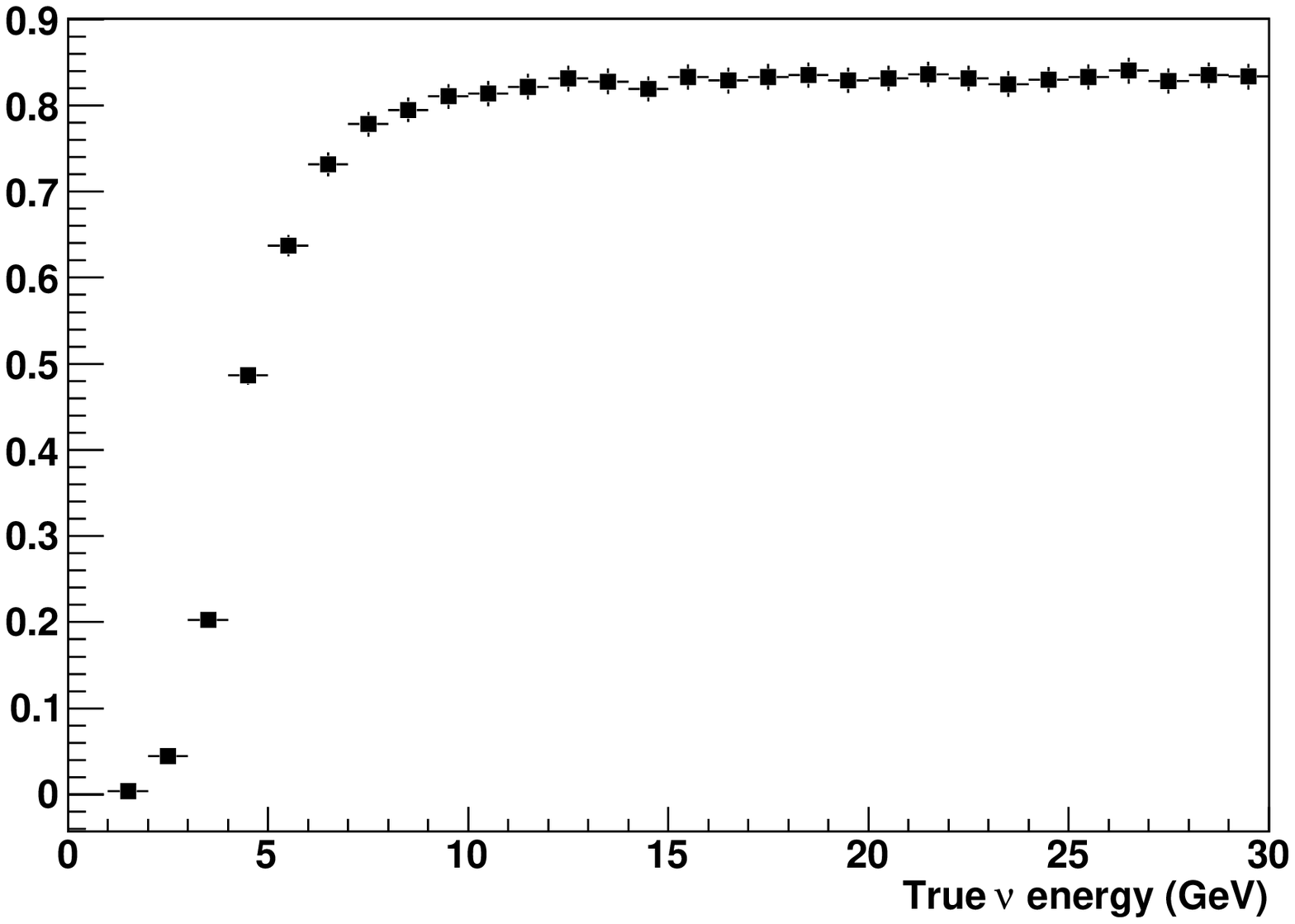} & 
      \includegraphics[scale=0.35]{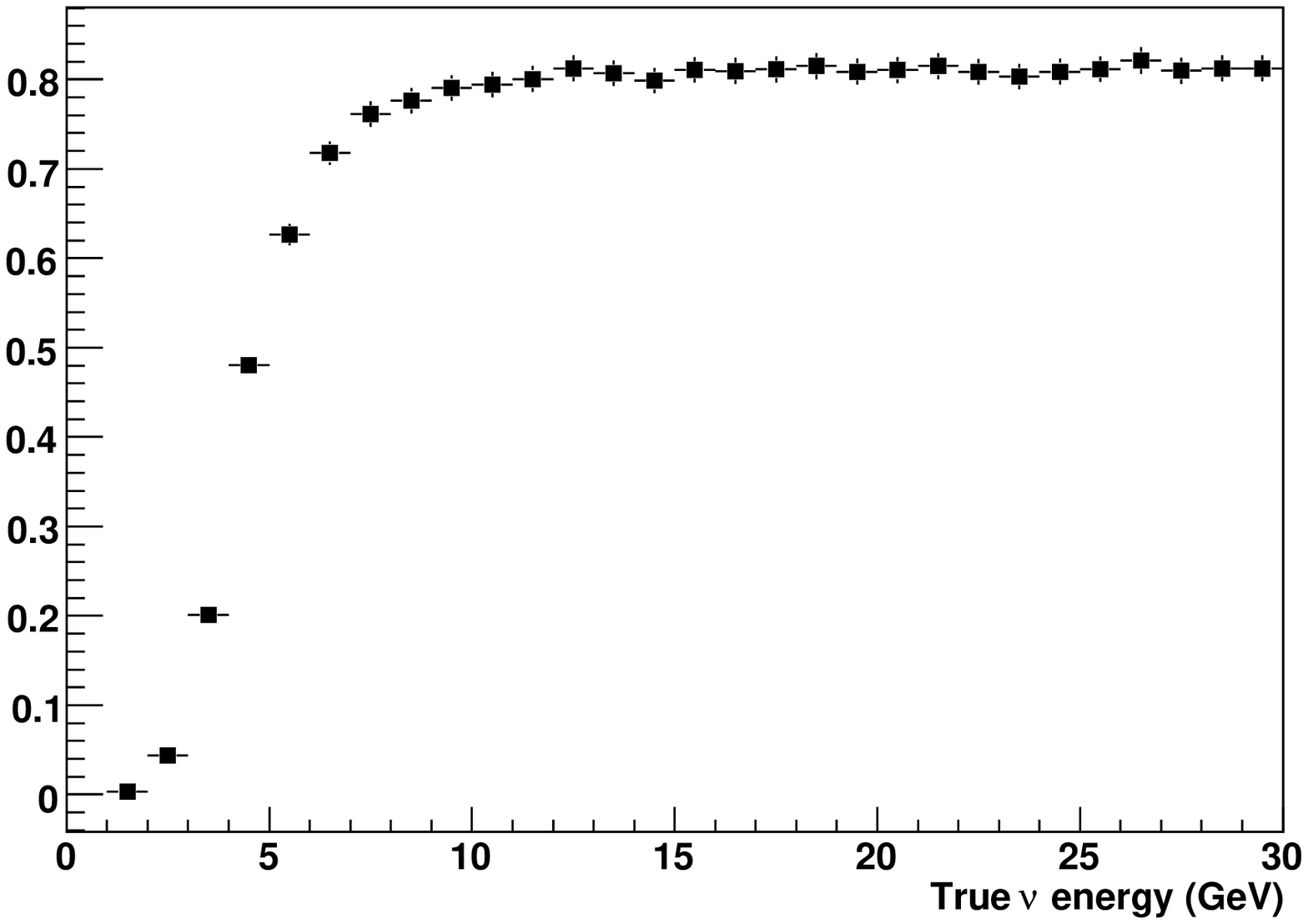}
    \end{array}$
    \includegraphics[scale=0.35]{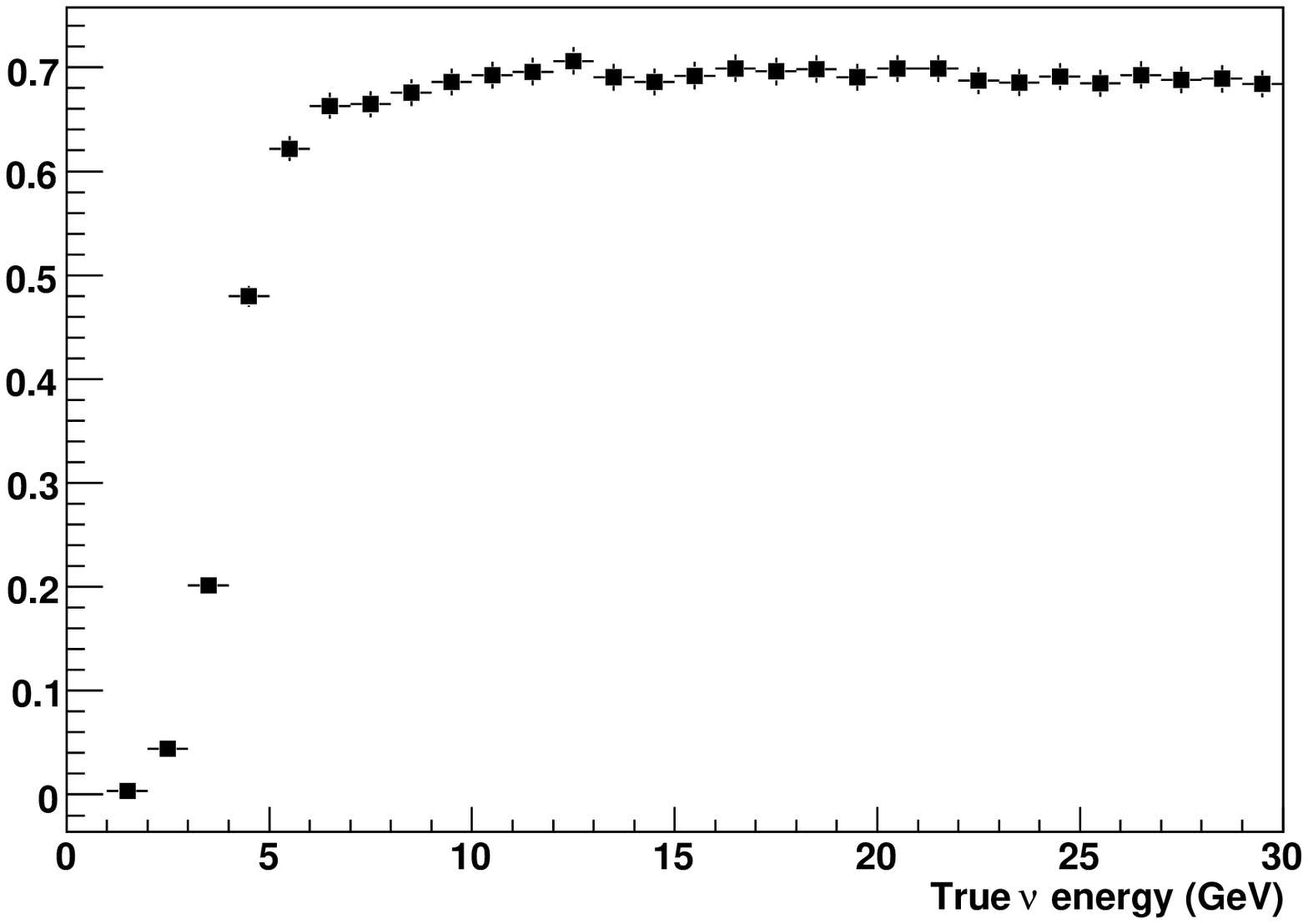} 
  \end{center}
  \caption{\emph{Expected signal identification efficiency: i) after track quality and $\nu_\mu$ CC selection cuts, ii) including fiducial cuts, and iii) including kinematic cuts.}}
  \label{fig:eff}
\end{figure}

Considering all types of events mentioned above and applying the most successful analysis chain described in table~\ref{tab:ana_cuts} the resulting signal efficiency and fractional backgrounds are those summarized in table~\ref{tab:backsum}. The evolution of the backgrounds for the different cuts is in Figs.~\ref{fig:numu_bkg}, ~\ref{fig:NC_bkg} and \ref{fig:nue_bkg}, while similar plots for the signal efficiency are those of Fig.~\ref{fig:eff}. While it is obvious that the effect of the kinematic cuts below 7~GeV true neutrino energy is small due to their application only to events reconstructed with energy greater than this value, it is important to remark that fiducial cuts do not affect the efficiency at low energies either. 

\begin{figure}
  \begin{center}$
    \begin{array}{cc}
      \includegraphics[scale=0.35]{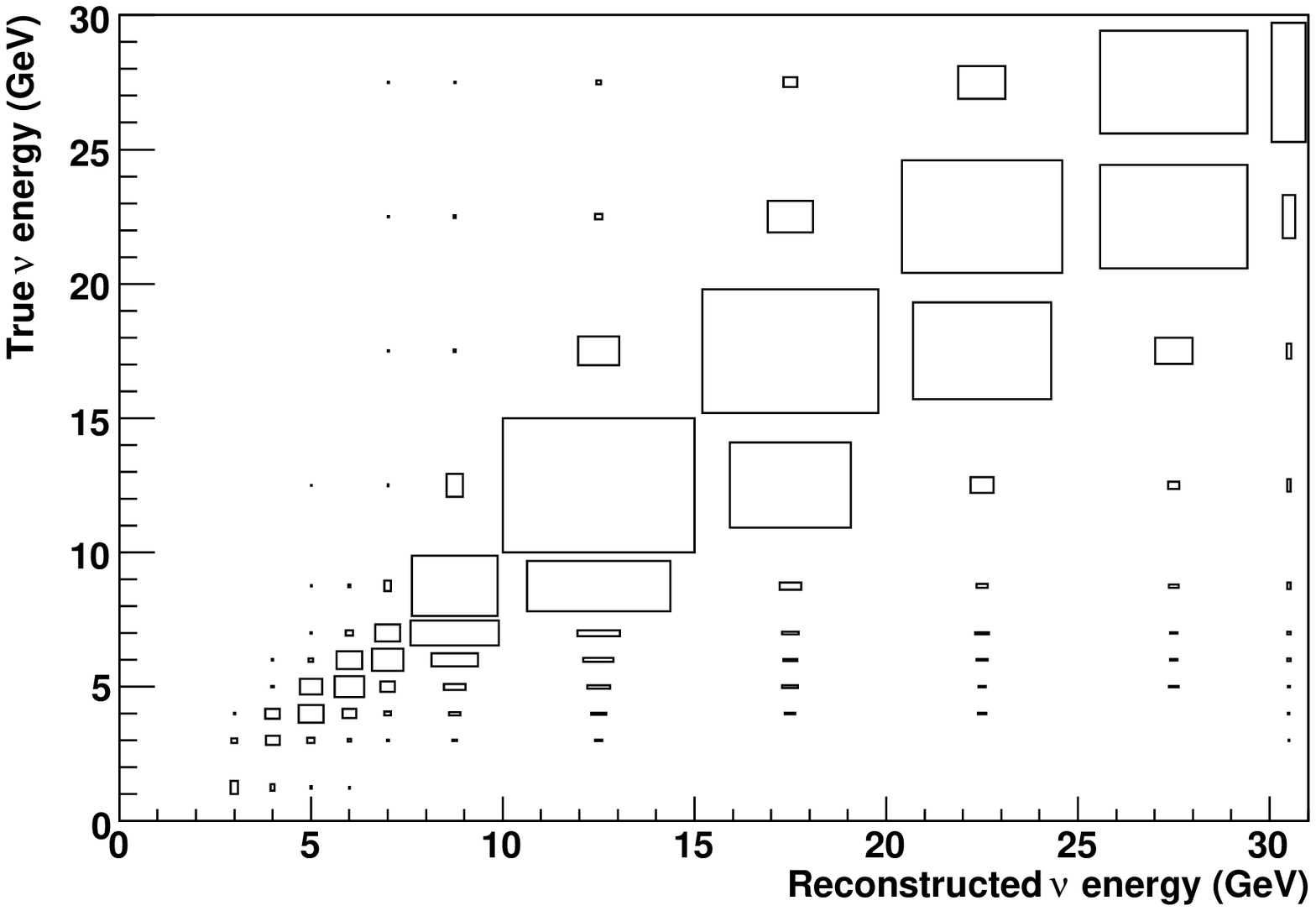} &
      \includegraphics[scale=0.35]{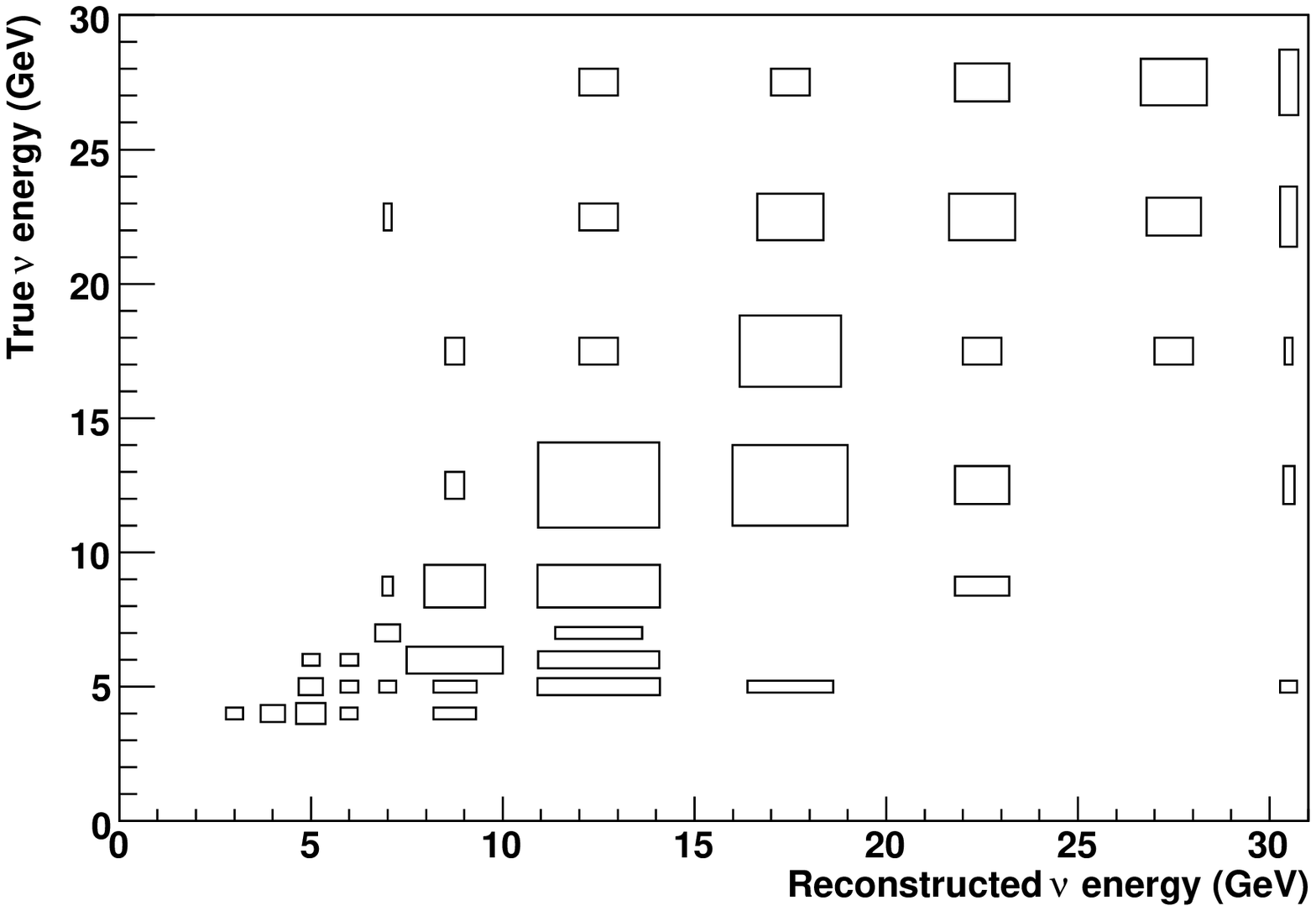} \\
      \includegraphics[scale=0.35]{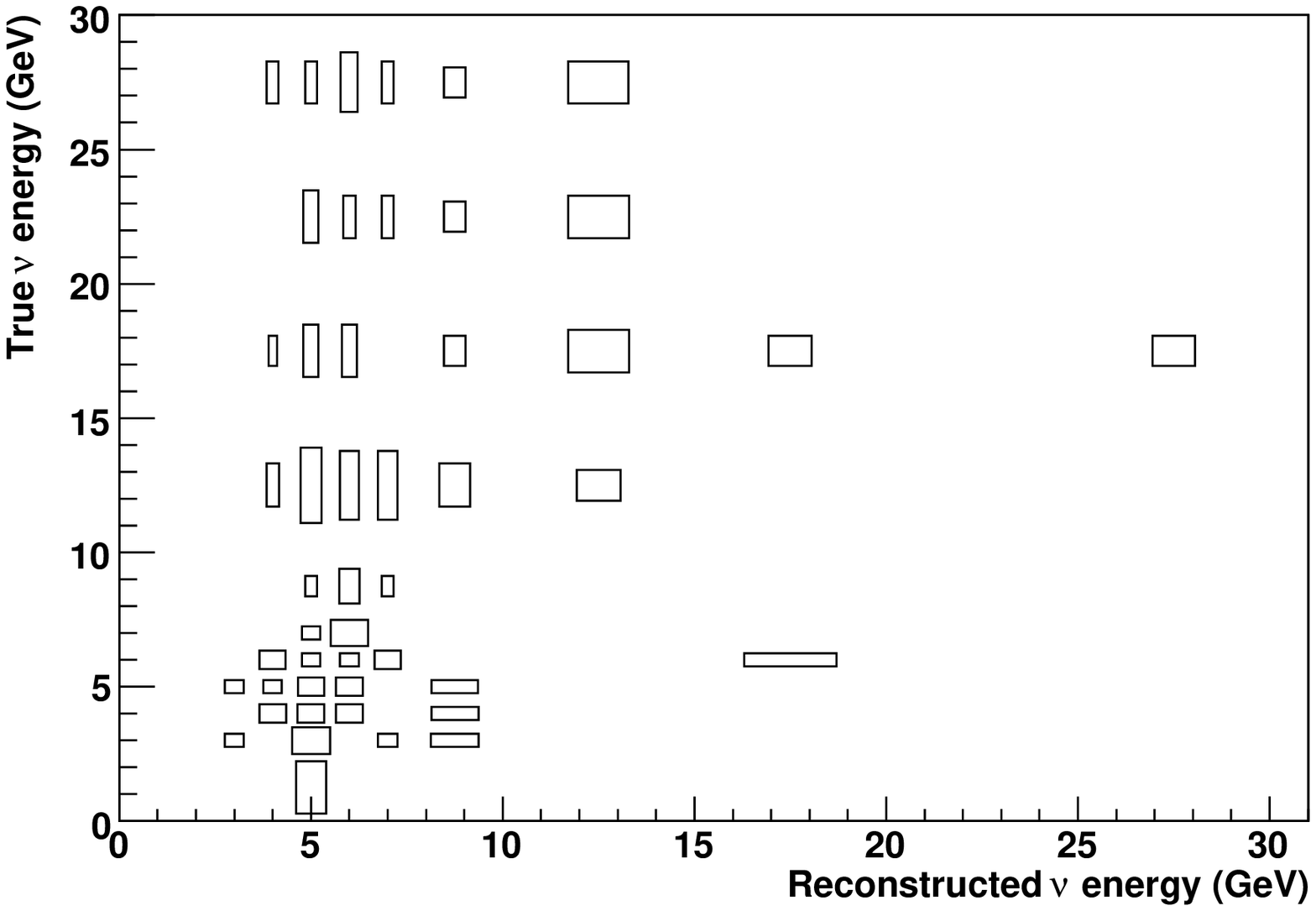} &
      \includegraphics[scale=0.35]{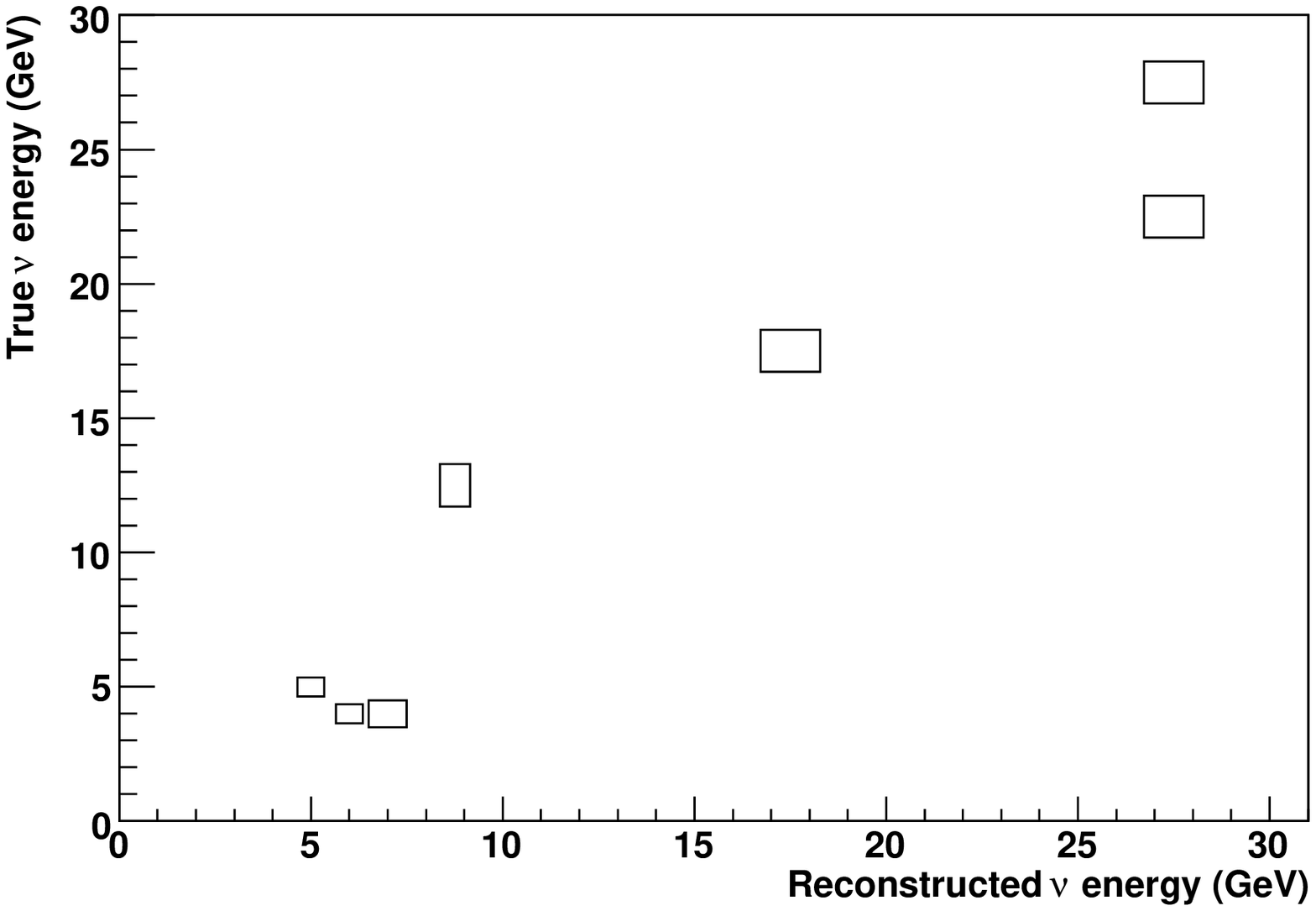} 
    \end{array}$
  \end{center}
  \caption{\emph{Response matrix in true/reconstructed neutrino energy for signal and backgrounds: i) signal efficiency, ii) backgrounds from $\nu_\mu$ CC, iii) NC and iv) $\nu_e$  CC.}}
  \label{fig:true_rec_matrix}
\end{figure}

Another important question is that of the relation between the true and the reconstructed neutrino energy for the different interaction types. The response matrices are shown graphically in Fig.~\ref{fig:true_rec_matrix} and numerically in App.~\ref{app:1}. 

%Background suppressions are at levels consistent with a successful analysis. 
%There are also a number of remaining events which could be rejected using a more 
%sophisticated kink finder capable of using both the slopes and position matching. 

%%Couple of examples??

%**********************************************************************
\section{Conclusions}
%**********************************************************************

\label{conc}
Through a combination of fiducial, track quality, $\nu_\mu$ CC selection and kinematical cuts, an analysis has been applied demonstrating the power of MIND to detect and identify $\nu_\mu$ CC DIS interactions in the presence of realistic reconstruction of the primary muon. The efficiency threshold currently lies between 3 and 4~GeV, 
and an efficiency plateau of 70\% is reached at about 6~GeV. While improved sensitivity could be achieved by lowering the threshold, as mentioned in~Sec.~\ref{MINpar}, this region would be dominated by quasi-elastic and resonance interactions which have not yet been considered. This type of interaction should contain less hadronic activity and thus the low energy pattern recognition efficiency should improve. An increased sample of successfully reconstructed events in this region should increase the efficiency of the golden channel analysis.

The powerful rejection afforded by the inclusion of the hadronic energy and direction vector highlights the importance of good hadronic reconstruction. While the hadronic energy has been reconstructed well in other similar experiments the direction vector requires careful consideration of both technology and analysis to achieve the required resolution. In a future publication we will include a re-optimisied MIND design, within a GEANT4 framework, where we will also take into account low energy quasi-elastic and resonance interactions. 

Compared to the baseline MIND presented in~\cite{Abe:2007bi}, where perfect pattern recognition was assumed, these new results show some improvement. The aforementioned study considered the charge and NC backgrounds and in both cases the results presented here are of similar level. The signal efficiency curve reaches a plateau at 70\% in the bin of $6-7$~GeV. The corresponding curve in the previous study reaches approximately the same level at a similar or slightly higher energy depending on the particular analysis. Using this efficiency curve, Neutrino Factory sensitivity studies were carried out in the context of the International Scoping Study (ISS) for a future neutrino facility \cite{Donini:2007rb}, demonstrating that a NF with two 50 ktonne MIND detectors at two different baselines has the largest $\theta_{13}-\delta_{CP}$ coverage out of all possible facilities. This paper reinforces the conclusions from that study by showing that the pattern recognition and reconstruction of events in MIND do not degrade the selection efficiency for the oscillated signal.
%need better explanation and probably cite.

%**********************************************************************
\section*{Acknowledgements}
%**********************************************************************
Thanks to the STFC and Ministerio de Ciencia e Innovaci\'on for support and to all collegues in both Glasgow and Valencia for all their help and advice.
\bibliographystyle{elsarticle-num}
\bibliography{MINDRef}
\newpage
\appendix
%will contain numeric response matrices.

%**********************************************************************
\section{Numeric summary of Response Matrices}
%**********************************************************************
\label{app:1}
This appendix summarises the response matrices of signal and all backgrounds in bins relevant to an oscillation analysis. In all tables columns represent the true neutrino energy in GeV and rows the reconstructed energy, also in GeV. The overflow bin in reconstructed energy represents all events with a reconstructed energy greater than the known maximum.
\newpage
\begin{table}
  \begin{tabular}{|l||c|c|c|c|c|c|c|c|c|c|c|c|}
    \hline
    & {\tiny 0-2.5} & {\tiny 2.5-3.5} & {\tiny 3.5-4.5} & {\tiny 4.5-5.5} & {\tiny 5.5-6.5} & {\tiny 6.5-7.5} & {\tiny 7.5-10} & {\tiny 10-15} & {\tiny 15-20} & {\tiny 20-25} & {\tiny 25-30} \\
    \hline
    \hline
{\tiny 0-2.5}  & {\scriptsize0} & {\scriptsize0} & {\scriptsize0} & {\scriptsize0} & {\scriptsize0} & {\scriptsize0} & {\scriptsize0} & {\scriptsize0} & {\scriptsize0} & {\scriptsize0} & {\scriptsize0} \\
	\hline
	    {\tiny 2.5-3.5}  & {\scriptsize1.78} & {\scriptsize1.26} & {\scriptsize0.01} & {\scriptsize0} & {\scriptsize0} & {\scriptsize0} & {\scriptsize0} & {\scriptsize0} & {\scriptsize0} & {\scriptsize0} & {\scriptsize0} \\
	    \hline
		{\tiny 3.5-4.5}  & {\scriptsize0.49} & {\scriptsize5.94} & {\scriptsize6.54} & {\scriptsize0.20} & {\scriptsize0.04} & {\scriptsize0} & {\scriptsize0} & {\scriptsize0} & {\scriptsize0} & {\scriptsize0} & {\scriptsize0} \\
\hline
{\tiny 4.5-5.5}  & {\scriptsize0.08} & {\scriptsize1.71} & {\scriptsize20.24} & {\scriptsize16.07} & {\scriptsize0.68} & {\scriptsize0.03} & {\scriptsize0.01} & {\scriptsize0} & {\scriptsize0} & {\scriptsize0} & {\scriptsize0} \\
\hline
{\tiny 5.5-6.5}  & {\scriptsize0.04} & {\scriptsize0.39} & {\scriptsize6.04} & {\scriptsize28.25} & {\scriptsize20.72} & {\scriptsize1.59} & {\scriptsize0.07} & {\scriptsize0} & {\scriptsize0} & {\scriptsize0} & {\scriptsize0} \\
\hline
{\tiny 6.5-7.5}  & {\scriptsize0} & {\scriptsize0.12} & {\scriptsize1.18} & {\scriptsize7.26} & {\scriptsize31.82} & {\scriptsize20.23} & {\scriptsize1.21} & {\scriptsize0.01} & {\scriptsize0} & {\scriptsize0} & {\scriptsize0} \\
\hline
{\tiny 7.5-10}  & {\scriptsize0} & {\scriptsize0.09} & {\scriptsize0.70} & {\scriptsize2.31} & {\scriptsize11.22} & {\scriptsize40.36} & {\scriptsize38.50} & {\scriptsize1.38} & {\scriptsize0.01} & {\scriptsize0.01} & {\scriptsize0.01} \\
\hline
{\tiny 10-15}  & {\scriptsize0} & {\scriptsize0.06} & {\scriptsize0.30} & {\scriptsize0.67} & {\scriptsize1.18} & {\scriptsize2.29} & {\scriptsize26.76} & {\scriptsize47.64} & {\scriptsize2.15} & {\scriptsize0.075} & {\scriptsize0.032} \\
\hline
{\tiny 15-20}  & {\scriptsize0} & {\scriptsize0} & {\scriptsize0.14} & {\scriptsize0.32} & {\scriptsize0.24} & {\scriptsize0.35} & {\scriptsize0.58} & {\scriptsize19.15} & {\scriptsize40.25} & {\scriptsize2.68} & {\scriptsize0.26} \\
\hline
{\tiny 20-25}  & {\scriptsize0} & {\scriptsize0} & {\scriptsize0.10} & {\scriptsize0.07} & {\scriptsize0.14} & {\scriptsize0.25} & {\scriptsize0.17} & {\scriptsize0.66} & {\scriptsize24.72} & {\scriptsize33.40} & {\scriptsize2.87} \\
\hline
{\tiny 25-30}  & {\scriptsize0} & {\scriptsize0} & {\scriptsize0} & {\scriptsize0.12} & {\scriptsize0.07} & {\scriptsize0.06} & {\scriptsize0.12} & {\scriptsize0.15} & {\scriptsize1.77} & {\scriptsize28.15} & {\scriptsize27.86} \\
\hline
{\tiny overflow}  & {\scriptsize0} & {\scriptsize0.01} & {\scriptsize0.04} & {\scriptsize0.09} & {\scriptsize0.33} & {\scriptsize0.40} & {\scriptsize0.44} & {\scriptsize0.43} & {\scriptsize0.62} & {\scriptsize4.90} & {\scriptsize37.72} \\
\hline
  \end{tabular}
  \caption{\emph{Signal Efficiency response matrix; All values $\times 10^{-2}$}}
  \label{sum:Eff}
\end{table}

\begin{table}
  \begin{tabular}{|l||c|c|c|c|c|c|c|c|c|c|c|c|}
    \hline
    & {\tiny 0-2.5} & {\tiny 2.5-3.5} & {\tiny 3.5-4.5} & {\tiny 4.5-5.5} & {\tiny 5.5-6.5} & {\tiny 6.5-7.5} & {\tiny 7.5-10} & {\tiny 10-15} & {\tiny 15-20} & {\tiny 20-25} & {\tiny 25-30} \\
    \hline
    \hline
{\tiny 0-2.5}  & {\scriptsize0} & {\scriptsize0} & {\scriptsize0} & {\scriptsize0} & {\scriptsize0} & {\scriptsize0} & {\scriptsize0} & {\scriptsize0} & {\scriptsize0} & {\scriptsize0} & {\scriptsize0} \\
\hline
{\tiny 2.5-3.5}  & {\scriptsize0} & {\scriptsize0} & {\scriptsize0.14} & {\scriptsize0} & {\scriptsize0} & {\scriptsize0} & {\scriptsize0} & {\scriptsize0} & {\scriptsize0} & {\scriptsize0} & {\scriptsize0} \\
\hline
{\tiny 3.5-4.5}  & {\scriptsize0} & {\scriptsize0} & {\scriptsize0.29} & {\scriptsize0} & {\scriptsize0} & {\scriptsize0} & {\scriptsize0} & {\scriptsize0} & {\scriptsize0} & {\scriptsize0} & {\scriptsize0} \\
\hline
{\tiny 4.5-5.5}  & {\scriptsize0} & {\scriptsize0} & {\scriptsize0.43} & {\scriptsize0.29} & {\scriptsize0.14} & {\scriptsize0} & {\scriptsize0} & {\scriptsize0} & {\scriptsize0} & {\scriptsize0} & {\scriptsize0} \\
\hline
{\tiny 5.5-6.5}  & {\scriptsize0} & {\scriptsize0} & {\scriptsize0.14} & {\scriptsize0.15} & {\scriptsize0.14} & {\scriptsize0} & {\scriptsize0} & {\scriptsize0} & {\scriptsize0} & {\scriptsize0} & {\scriptsize0} \\
\hline
{\tiny 6.5-7.5}  & {\scriptsize0} & {\scriptsize0} & {\scriptsize0} & {\scriptsize0.15} & {\scriptsize0} & {\scriptsize0.29} & {\scriptsize0.06} & {\scriptsize0} & {\scriptsize0} & {\scriptsize0.03} & {\scriptsize0} \\
\hline
{\tiny 7.5-10}  & {\scriptsize0} & {\scriptsize0} & {\scriptsize0.14} & {\scriptsize0.15} & {\scriptsize0.72} & {\scriptsize0} & {\scriptsize0.29} & {\scriptsize0.03} & {\scriptsize0.03} & {\scriptsize0} & {\scriptsize0} \\
\hline
{\tiny 10-15}  & {\scriptsize0} & {\scriptsize0} & {\scriptsize0} & {\scriptsize0.29} & {\scriptsize0.29} & {\scriptsize0.15} & {\scriptsize0.29} & {\scriptsize0.29} & {\scriptsize0.03} & {\scriptsize0.03} & {\scriptsize0.03} \\
\hline
{\tiny 15-20}  & {\scriptsize0} & {\scriptsize0} & {\scriptsize0} & {\scriptsize0.15} & {\scriptsize0} & {\scriptsize0} & {\scriptsize0} & {\scriptsize0.26} & {\scriptsize0.20} & {\scriptsize0.09} & {\scriptsize0.03} \\
\hline
{\tiny 20-25}  & {\scriptsize0} & {\scriptsize0} & {\scriptsize0} & {\scriptsize0} & {\scriptsize0} & {\scriptsize0} & {\scriptsize0.06} & {\scriptsize0.06} & {\scriptsize0.03} & {\scriptsize0.09} & {\scriptsize0.06} \\
\hline
{\tiny 25-30}  & {\scriptsize0} & {\scriptsize0} & {\scriptsize0} & {\scriptsize0} & {\scriptsize0} & {\scriptsize0} & {\scriptsize0} & {\scriptsize0} & {\scriptsize0.03} & {\scriptsize0.06} & {\scriptsize0.09} \\
\hline
{\tiny overflow}  & {\scriptsize0} & {\scriptsize0} & {\scriptsize0} & {\scriptsize0.15} & {\scriptsize0} & {\scriptsize0} & {\scriptsize0} & {\scriptsize0.06} & {\scriptsize0.03} & {\scriptsize0.14} & {\scriptsize0.17} \\
\hline
  \end{tabular}
  \caption{\emph{$\overline{\nu}_\mu$ CC background response matrix; All values $\times 10^{-3}$}}
  \label{sum:char}
\end{table}

\begin{table}
  \begin{tabular}{|l||c|c|c|c|c|c|c|c|c|c|c|c|}
    \hline
    & {\tiny 0-2.5} & {\tiny 2.5-3.5} & {\tiny 3.5-4.5} & {\tiny 4.5-5.5} & {\tiny 5.5-6.5} & {\tiny 6.5-7.5} & {\tiny 7.5-10} & {\tiny 10-15} & {\tiny 15-20} & {\tiny 20-25} & {\tiny 25-30} \\
    \hline
    \hline
{\tiny 0-2.5}  & {\scriptsize0} & {\scriptsize0} & {\scriptsize0} & {\scriptsize0} & {\scriptsize0} & {\scriptsize0} & {\scriptsize0} & {\scriptsize0} & {\scriptsize0} & {\scriptsize0} & {\scriptsize0} \\
\hline
{\tiny 2.5-3.5}  & {\scriptsize0} & {\scriptsize0.01} & {\scriptsize0} & {\scriptsize0.01} & {\scriptsize0} & {\scriptsize0} & {\scriptsize0} & {\scriptsize0} & {\scriptsize0} & {\scriptsize0} & {\scriptsize0} \\
\hline
{\tiny 3.5-4.5}  & {\scriptsize0} & {\scriptsize0} & {\scriptsize0.02} & {\scriptsize0.01} & {\scriptsize0.02} & {\scriptsize0} & {\scriptsize0} & {\scriptsize0} & {\scriptsize0} & {\scriptsize0} & {\scriptsize0} \\
\hline
{\tiny 4.5-5.5}  & {\scriptsize0.03} & {\scriptsize0.05} & {\scriptsize0.02} & {\scriptsize0.02} & {\scriptsize0.01} & {\scriptsize0.01} & {\scriptsize0} & {\scriptsize0.01} & {\scriptsize0.01} & {\scriptsize0.01} & {\scriptsize0} \\
\hline
{\tiny 5.5-6.5}  & {\scriptsize0} & {\scriptsize0} & {\scriptsize0.02} & {\scriptsize0.02} & {\scriptsize0.01} & {\scriptsize0.04} & {\scriptsize0.01} & {\scriptsize0.01} & {\scriptsize0.01} & {\scriptsize0} & {\scriptsize0.01} \\
\hline
{\tiny 6.5-7.5}  & {\scriptsize0} & {\scriptsize0.01} & {\scriptsize0} & {\scriptsize0} & {\scriptsize0.02} & {\scriptsize0} & {\scriptsize0} & {\scriptsize0.01} & {\scriptsize0} & {\scriptsize0.01} & {\scriptsize0} \\
\hline
{\tiny 7.5-10}  & {\scriptsize0} & {\scriptsize0.01} & {\scriptsize0.01} & {\scriptsize0.01} & {\scriptsize0} & {\scriptsize0} & {\scriptsize0} & {\scriptsize0.01} & {\scriptsize0} & {\scriptsize0} & {\scriptsize0} \\
\hline
{\tiny 10-15}  & {\scriptsize0} & {\scriptsize0} & {\scriptsize0} & {\scriptsize0} & {\scriptsize0} & {\scriptsize0} & {\scriptsize0} & {\scriptsize0} & {\scriptsize0.01} & {\scriptsize0.01} & {\scriptsize0} \\
\hline
{\tiny 15-20}  & {\scriptsize0} & {\scriptsize0} & {\scriptsize0} & {\scriptsize0} & {\scriptsize0.01} & {\scriptsize0} & {\scriptsize0} & {\scriptsize0} & {\scriptsize0} & {\scriptsize0} & {\scriptsize0} \\
\hline
{\tiny 20-25}  & {\scriptsize0} & {\scriptsize0} & {\scriptsize0} & {\scriptsize0} & {\scriptsize0} & {\scriptsize0} & {\scriptsize0} & {\scriptsize0} & {\scriptsize0} & {\scriptsize0} & {\scriptsize0} \\
\hline
{\tiny 25-30}  & {\scriptsize0} & {\scriptsize0} & {\scriptsize0} & {\scriptsize0} & {\scriptsize0} & {\scriptsize0} & {\scriptsize0} & {\scriptsize0} & {\scriptsize0} & {\scriptsize0} & {\scriptsize0} \\
\hline
{\tiny overflow}  & {\scriptsize0} & {\scriptsize0} & {\scriptsize0} & {\scriptsize0} & {\scriptsize0} & {\scriptsize0} & {\scriptsize0} & {\scriptsize0} & {\scriptsize0} & {\scriptsize0} & {\scriptsize0} \\
\hline
  \end{tabular}
  \caption{\emph{Neutral current background response matrix; All values $\times 10^{-3}$}}
  \label{sum:NC}
\end{table}

\begin{table}
  \begin{tabular}{|l||c|c|c|c|c|c|c|c|c|c|c|c|}
    \hline
    & {\tiny 0-2.5} & {\tiny 2.5-3.5} & {\tiny 3.5-4.5} & {\tiny 4.5-5.5} & {\tiny 5.5-6.5} & {\tiny 6.5-7.5} & {\tiny 7.5-10} & {\tiny 10-15} & {\tiny 15-20} & {\tiny 20-25} & {\tiny 25-30} \\
    \hline
    \hline
{\tiny 0-2.5}  & {\scriptsize0} & {\scriptsize0} & {\scriptsize0} & {\scriptsize0} & {\scriptsize0} & {\scriptsize0} & {\scriptsize0} & {\scriptsize0} & {\scriptsize0} & {\scriptsize0} & {\scriptsize0} \\
\hline
{\tiny 2.5-3.5}  & {\scriptsize0} & {\scriptsize0} & {\scriptsize0} & {\scriptsize0} & {\scriptsize0} & {\scriptsize0} & {\scriptsize0} & {\scriptsize0} & {\scriptsize0} & {\scriptsize0} & {\scriptsize0} \\
\hline
{\tiny 3.5-4.5}  & {\scriptsize0} & {\scriptsize0} & {\scriptsize0} & {\scriptsize0} & {\scriptsize0} & {\scriptsize0} & {\scriptsize0} & {\scriptsize0} & {\scriptsize0} & {\scriptsize0} & {\scriptsize0} \\
\hline
{\tiny 4.5-5.5}  & {\scriptsize0} & {\scriptsize0} & {\scriptsize0} & {\scriptsize0.03} & {\scriptsize0} & {\scriptsize0} & {\scriptsize0} & {\scriptsize0} & {\scriptsize0} & {\scriptsize0} & {\scriptsize0} \\
\hline
{\tiny 5.5-6.5}  & {\scriptsize0} & {\scriptsize0} & {\scriptsize0.03} & {\scriptsize0} & {\scriptsize0} & {\scriptsize0} & {\scriptsize0} & {\scriptsize0} & {\scriptsize0} & {\scriptsize0} & {\scriptsize0} \\
\hline
{\tiny 6.5-7.5}  & {\scriptsize0} & {\scriptsize0} & {\scriptsize0.06} & {\scriptsize0} & {\scriptsize0} & {\scriptsize0} & {\scriptsize0} & {\scriptsize0} & {\scriptsize0} & {\scriptsize0} & {\scriptsize0} \\
\hline
{\tiny 7.5-10}  & {\scriptsize0} & {\scriptsize0} & {\scriptsize0} & {\scriptsize0} & {\scriptsize0} & {\scriptsize0} & {\scriptsize0} & {\scriptsize0.01} & {\scriptsize0} & {\scriptsize0} & {\scriptsize0} \\
\hline
{\tiny 10-15}  & {\scriptsize0} & {\scriptsize0} & {\scriptsize0} & {\scriptsize0} & {\scriptsize0} & {\scriptsize0} & {\scriptsize0} & {\scriptsize0} & {\scriptsize0} & {\scriptsize0} & {\scriptsize0} \\
\hline
{\tiny 15-20}  & {\scriptsize0} & {\scriptsize0} & {\scriptsize0} & {\scriptsize0} & {\scriptsize0} & {\scriptsize0} & {\scriptsize0} & {\scriptsize0} & {\scriptsize0.01} & {\scriptsize0} & {\scriptsize0} \\
\hline
{\tiny 20-25}  & {\scriptsize0} & {\scriptsize0} & {\scriptsize0} & {\scriptsize0} & {\scriptsize0} & {\scriptsize0} & {\scriptsize0} & {\scriptsize0} & {\scriptsize0} & {\scriptsize0} & {\scriptsize0} \\
\hline
{\tiny 25-30}  & {\scriptsize0} & {\scriptsize0} & {\scriptsize0} & {\scriptsize0} & {\scriptsize0} & {\scriptsize0} & {\scriptsize0} & {\scriptsize0} & {\scriptsize0} & {\scriptsize0.01} & {\scriptsize0.01} \\
\hline
{\tiny overflow}  & {\scriptsize0} & {\scriptsize0} & {\scriptsize0} & {\scriptsize0} & {\scriptsize0} & {\scriptsize0} & {\scriptsize0} & {\scriptsize0} & {\scriptsize0} & {\scriptsize0} & {\scriptsize0} \\
\hline
  \end{tabular}
  \caption{\emph{$\nu_e$ CC background response matrix; All values $\times 10^{-3}$}}
  \label{sum:elec}
\end{table}

%% \begin{thebibliography}{00}


%% \bibitem{}

%% \end{thebibliography}

\end{document}